\documentclass{revtex4}
\usepackage{graphicx}
\usepackage{amsmath}
\usepackage{amsfonts}
\usepackage{amssymb}
\usepackage{subfigure}

\begin{document}

\title{The interaction of Airy waves and solitons in the three-wave system}
\author{Thawatchai Mayteevarunyoo$^{1}$ and Boris A. Malomed$^{2,3}$}
\address{$^{1}$Department of Telecommunication Engineering, Mahanakorn University
of Technology, Bangkok 10530, Thailand\\
$^{2}$Department of Physical Electronics, School of Electrical Engineering,
Faculty of Engineering, Tel Aviv University, Tel Aviv 69978, Israel\\
$^{3}$Laboratory of Nonlinear-Optical Informatics, ITMO University, St. Petersburg
197101, Russia}

\begin{abstract}
We employ the generic three-wave system, with the $\chi ^{(2)}$ interaction
between two components of the fundamental-frequency (FF) wave and
second-harmonic (SH) one, to consider collisions of truncated Airy waves
(TAWs) and three-wave solitons in a setting which is not available in other
nonlinear systems. The advantage is that the single-wave TAWs, carried by
either one of the FF component, are not distorted by the nonlinearity and
are stable, three-wave solitons being stable too in the same system. The
collision between mutually symmetric TAWs, carried by the different FF
components, transforms them into a set of solitons, the number of which
decreases with the increase of the total power. The TAW absorbs an incident
small-power soliton, and a high-power soliton absorbs the TAW. Between these
limits, the collision with an incident soliton converts the TAW into two
solitons, with a remnant of the TAW attached to one of them, or leads to
formation of a complex TAW-soliton bound state. At large velocities, the
collisions become quasi-elastic.
\end{abstract}

\maketitle

\section{Introduction}

Airy waves represent a universal transmission mode in linear media governed
by linear Schr\"{o}dinger equations, which makes it possible to send free
beams along self-bending trajectories \cite{Berry}. While the full Airy mode
has a divergent total norm, it was predicted that its finite-norm version,
i.e., a \textit{truncated Airy wave}, TAW, supports similar curvilinear
transmission \cite{Christo}. The propagation of TAWs was demonstrated,
theoretically and experimentally, in various settings in optics \cite%
{Christo}-\cite{Efremidis}, plasmonics \cite{plasm1}-\cite{plasm3}, electron
beams \cite{el}, matter waves \cite{BEC}, gas discharge \cite{discharge},
and water waves \cite{water}.

While the Airy waves and their truncated version are eigenmodes of the
linear transmission, the propagation of their counterparts distorted by the
self-interaction was also studied in media with cubic \cite{ChrSegev-nonlin}-%
\cite{Conti} and quadratic (alias $\chi ^{(2)}$) \cite{Ady-3wave}-\cite%
{Ady-Moti} nonlinearities. In the latter case, the TAW launched as the
fundamental-frequency (FF) wave gives rise to two-color beams, generating
the second-harmonic (SH) component via the $\chi ^{(2)}$-mediated
upconversion.

Because TAWs feature multi-lobe structures \cite{Christo}, while stable
eigenmodes of nonlinear media are single-hump solitons, a natural outcome of
the nonlinear transformation may be \textit{shedding of solitons} by the TAW
under the action of the cubic nonlinearity \cite{Marom,water}. A somewhat
similar outcome of the evolution is downconversion of a TAW launched, as an
exact linear SH mode, into a set of $\chi ^{(2)}$ solitons, initiated by the
parametric instability of the SH wave seeded by FF perturbations, in one-
\cite{we1} and two-dimensional \cite{we2} settings alike.

Another natural possibility in nonlinear media is interaction between TAWs,
as well as between them and solitons. Such interactions were theoretically
studied in cubic \cite{Marom2}-\cite{Milivoj}, \cite{Radik-VVK} and
photorefractive \cite{photorefr} media, in systems with nonlocal
nonlinearities \cite{nonlocal,nonlocal2}, and in a model of coherent atomic
media \cite{coherent}. In all these cases, the situation is rather tricky,
as the nonlinearity distorts the TAW even without its interaction with
another mode. In the present work, we aim to propose a system which makes it
possible to study interactions between \emph{undistorted} stable Airy waves,
as well as their interactions with stable solitons. This model is based on
the nondegenerate, i.e., three-wave, system with quadratic couplings between
two FF components and the SH wave (unlike the degenerate version, in which
the $\chi ^{(2)}$ terms couple the SH to the single FF component, that was
considered in previous works dealing with Airy waves in quadratic media \cite%
{Ady-3wave}-\cite{Ady-Moti}). The two FF complex wave fields, $u$ and $v$,
usually represent orthogonal polarizations of the waves carried by the FF,
while the SH field, $w$, is represented by a single polarization. In the
spatial domain, the three-wave system, which corresponds to the so-called
Type-II $\chi ^{(2)}$ interaction in a planar waveguide, is modeled by the
well-known propagation equations, written in the scaled form corresponding
to the paraxial approximation \cite{rev1}-\cite{rev3}:
\begin{eqnarray}
iu_{z}+bu+\frac{1}{2}u_{xx}+v^{\ast }w &=&0,  \notag \\
iv_{z}-bv+\frac{1}{2}v_{xx}+u^{\ast }w &=&0,  \label{type-II} \\
2iw_{z}-qw+\frac{1}{2}w_{xx}+uv &=&0,  \notag
\end{eqnarray}%
where $z$ and $x$ are the propagation distance and transverse coordinate,
real coefficients $b$ and $q$ represent the birefringence and mismatch
parameters, respectively, and $\ast $ stands for the complex conjugate
field. The system conserves the total power (alias the Manley-Rowe
invariant, or total norm),
\begin{equation}
P=\int_{-\infty }^{+\infty }\left( |u|^{2}+|v|^{2}+4|w|^{2}\right) dx\equiv
P_{u}+P_{v}+P_{w},  \label{P1D}
\end{equation}%
the total momentum,%
\begin{equation}
M=i\int_{-\infty }^{+\infty }\left( uu_{x}^{\ast }+vv_{x}^{\ast
}+2ww_{x}^{\ast }\right) dx,  \label{M1D}
\end{equation}%
and the Hamiltonian,%
\begin{equation}
H=\int_{-\infty }^{+\infty }\left(
|u_{x}|^{2}+|v_{x}|^{2}+|w_{x}|^{2}+b|v|^{2}-|u|^{2}+q|w|^{2}+u^{\ast
}v^{\ast }w+uvw^{\ast }\right) dx.
\end{equation}

Scaled variables $x$ and $z$ are measured, respectively, in units of the
characteristic size of $\chi ^{(2)}$ patterns (such as solitons), which is $%
l\sim 50~\mathrm{\mu }$m \cite{rev1}-\cite{rev3}, and the respective
diffraction length, $\Delta z\sim \left( 2\pi /\lambda \right) l^{2}\sim 1$
cm, for the carrier wavelength $\lambda \sim 1$ $\mathrm{\mu }$m. Then, $q=1$
and $b=1$ correspond, in physical units, to the mismatch $1$ cm$^{-1}$. As
concerns tilted (``moving") solitons (e.g., ones with slopes
$1/4$ and $1/3$ in the scaled variables, which are shown below in Fig. \ref%
{fig3}(a) and \ref{fig8}(a)-\ref{fig_additional}(a), respectively), their
slope, as measured in real coordinates, is smaller by a factor $\sim 0.01$.

Actually, parameter $b$ can be eliminated from Eqs. (\ref{type-II}) by means
of a substitution,%
\begin{equation}
\left\{ u\left( x,z\right) ,v\left( x,z\right) \right\} \equiv \left\{
e^{ibz}\tilde{u}\left( x,z\right) ,e^{-ibz}\tilde{v}\left( x,z\right)
\right\} .  \label{uv}
\end{equation}%
On the other hand, as concerns soliton solutions considered below, fixing
propagation constants of the $u$ and $v$ waves and varying $b$, it is
possible to analyze the change of the solutions following the variation of
the difference between full propagation constants of the two FF components,
see Fig. \ref{fig2}(a) below.

An advantage offered by system (\ref{type-II}) is that it admits exact\emph{%
\ stable} TAW solutions, with $v=w=0$ and the same form of the $u$ component
as in the single linear Schr\"{o}dinger equation \cite{Christo}:%
\begin{gather}
u_{\mathrm{TAW}}\left( x,z\right) =u_{0}\mathrm{Ai}\left( \alpha x-\frac{%
\alpha ^{4}}{4}z^{2}+i\aleph \alpha ^{2}z\right)  \notag \\
\times \exp \left( -\frac{i}{12}\alpha ^{6}z^{3}+\frac{i}{2}\alpha
^{3}xz\right)  \notag \\
\times \exp \left( \aleph \alpha x-\frac{1}{2}\aleph \alpha ^{4}z^{2}+\frac{i%
}{2}\aleph ^{2}\alpha ^{2}z+ibz\right) ,  \label{1D-AW}
\end{gather}%
where $\mathrm{Ai}$ is the standard Airy function, while $u_{0}$, $\alpha $,
and $\aleph $ are positive constants that define, respectively, the
amplitude, intrinsic scale, and truncation of the Airy wave, as seen from
the form of the initial condition which generates the TAW solution:%
\begin{equation}
u\left( x\right) |_{z=0}=u_{0}\mathrm{Ai}\left( \alpha x\right) \exp \left(
\aleph \alpha x\right) ,~v\left( x\right) |_{z=0}=w\left( x\right) |_{z=0}=0.
\label{z=0}
\end{equation}%
The total momentum (\ref{M1D}) of real waveform (\ref{z=0}) is zero, while
its total power is \cite{Christo}%
\begin{equation}
P_{\mathrm{TAW}}=\frac{u_{0}^{2}}{\sqrt{8\pi \aleph }\alpha }\exp \left(
\frac{2}{3}\aleph ^{3}\right) .  \label{P1DAW}
\end{equation}

A symmetric counterpart of the TAW, with the center shifted by $x_{0}$ and
inverted orientation (which is necessary for the consideration of collisions
between the TAWs, see below) can be obtained from Eq. (\ref{1D-AW}) by
setting%
\begin{equation}
v_{\mathrm{TAW}}\left( x,z\right) =u_{\mathrm{TAW}}\left(
x-x_{0},z,b\rightarrow -b,\alpha \rightarrow -\alpha \right) ,~u=w=0
\label{tilde}
\end{equation}%
(note that this transformation does not change the sign of truncation
parameter $\aleph $).

In previous work \cite{we1}, TAW was taken as the exact solution of the
linear version of the equation for the SH field, $w$, in system (\ref%
{type-II}), while setting $u=v=0$. The so introduced TAW was subject to the
\textit{parametric instability} seeded by small perturbations in the $u$ and
$v$ fields, which leads to spontaneous downconversion of the TAW into a
cluster of $\chi ^{(2)}$ solitons and additional radiation jets (in fact,
the degenerate two-wave version of the $\chi ^{(2)}$ system was considered
in that case, but the TAW state in the SH field remains unstable in the
three-wave system as well). In the present case, it is easy to demonstrate,
considering small perturbations in the $v$ and $w$ fields, that the exact
TAW solution (\ref{1D-AW}) in the $u$ field, as well as its $v$-field
counterpart, given by Eq. (\ref{tilde}), are \emph{stable} against
perturbations (this conclusion is corroborated by direct simulations of the
perturbed evolution of the TAWs, which are not displayed here, as they do
not show anything essentially new). Thus, the system (\ref{type-II})
suggests a novel possibility to consider nonlinear interaction between two
stable undistorted TAWs, namely, ones given by Eqs. (\ref{1D-AW}) and (\ref%
{tilde}). Further, the three-wave system gives rise to three-component
solitons \cite{Goth,Boardman}. The Galilean invariance of the system makes
it possible to boost the solitons to an arbitrary velocity (actually, moving
solitons are tilted spatial beams), hence the same system makes it possible
to study collisions of the three-component solitons with the stable TAW
created in one FF component, as well as soliton-soliton collisions.

The rest of the paper is organized as follows. The three-wave solitons of
system (\ref{type-II}), which are available in a particular analytical form,
and in a numerical form in the general case, are introduced in Section II.
Typical examples of elastic and inelastic collisions between moving stable
solitons are presented in Section II too. Collisions between the TAW and its
mirror-image counterpart, given \ by Eqs. (\ref{1D-AW}) and (\ref{tilde}),
are considered by means of direct simulations in Section III. The analysis
is continued in Section IV, which reports results of systematic simulations
of collisions between moving three-wave solitons and TAWs. The paper is
concluded by Section V.

\section{Three-component solitons}

Soliton solutions to Eq. (\ref{type-II}) are looked for as%
\begin{equation}
u(x,z)=U_{\mathrm{sol}}(\xi )e^{ik_{1}z+icx},~v(x,z)=V_{\mathrm{sol}}(\xi
)e^{ik_{2}z+icx},~w(x,z)=W_{\mathrm{sol}}(\xi )e^{i\left( k_{1}+k_{2}\right)
z+2icx},  \label{sol}
\end{equation}%
where $\xi \equiv x-cz$, $c$ is the soliton's velocity (actually, the tilt
in the spatial domain), $k_{1}$ and $k_{2}$ are two independent propagation
constants, which are free parameters of the soliton family, and real
functions $U_{\mathrm{sol}}$, $V_{\mathrm{sol}}$, and $W_{\mathrm{sol}}$ are
solutions of the stationary equations:%
\begin{gather}
\frac{1}{2}\frac{d^{2}U}{d\xi ^{2}}+VW=\left( k_{1}-b\right) U,  \notag \\
\frac{1}{2}\frac{d^{2}V}{d\xi ^{2}}+UW=\left( k_{2}+b\right) V,  \label{UVW}
\\
\frac{1}{2}\frac{d^{2}W}{d\xi ^{2}}+UV=\left[ 2\left( k_{1}+k_{2}\right) +q%
\right] W.  \notag
\end{gather}%
Obvious conditions necessary for the existence of exponentially localized
solutions to these equations are%
\begin{equation}
-k_{2}<b<k_{1},~k_{1}+k_{2}>-q/2.~  \label{<<>}
\end{equation}

Following Ref. \cite{Goth}, it is straightforward to find particular exact
solutions of Eq. (\ref{UVW}), with mutually symmetric $U$ and $V$
components, for $q<0$:%
\begin{equation}
k_{1}=b-\left( q/3\right) ,~k_{2}=-b-\left( q/3\right) ,  \label{k12}
\end{equation}%
\begin{equation}
U(x)=V(x)=-W(x)=\left( q/2\right) \mathrm{sech}^{2}\left( \sqrt{-q/6}\xi
\right) .  \label{exact}
\end{equation}%
The total power (\ref{P1D}) and momentum (\ref{M1D}) of the exact soliton
solutions given by Eqs. (\ref{sol}) and (\ref{exact}) are%
\begin{equation}
P=2\sqrt{-6q^{3}}  \label{Psol}
\end{equation}%
$~$ and $M=cP$, the latter relation between $M$ and $P$ being valid for all
solitons (not only for exact ones given by Eqs. (\ref{k12}) and (\ref{exact}%
)).

Symmetric solitons with $U=V$ are tantamount to their well-known
counterparts in the degenerate (two-component) version of the $\chi ^{(2)}$
system; in particular, exact solutions given by Eqs. (\ref{k12}) and (\ref%
{exact}) are equivalent to the Karamzin-Sukhorukov soliton solutions of the
degenerate system \cite{rev1}-\cite{rev3}. However, stability of the
symmetric solitons may be different in the three-wave system -- in
particular, three-wave solitons with the symmetry between $U$ and $V$ should
be stable against symmetry-breaking perturbations.

\begin{figure}[tbp]
\centering\includegraphics[width=3in]{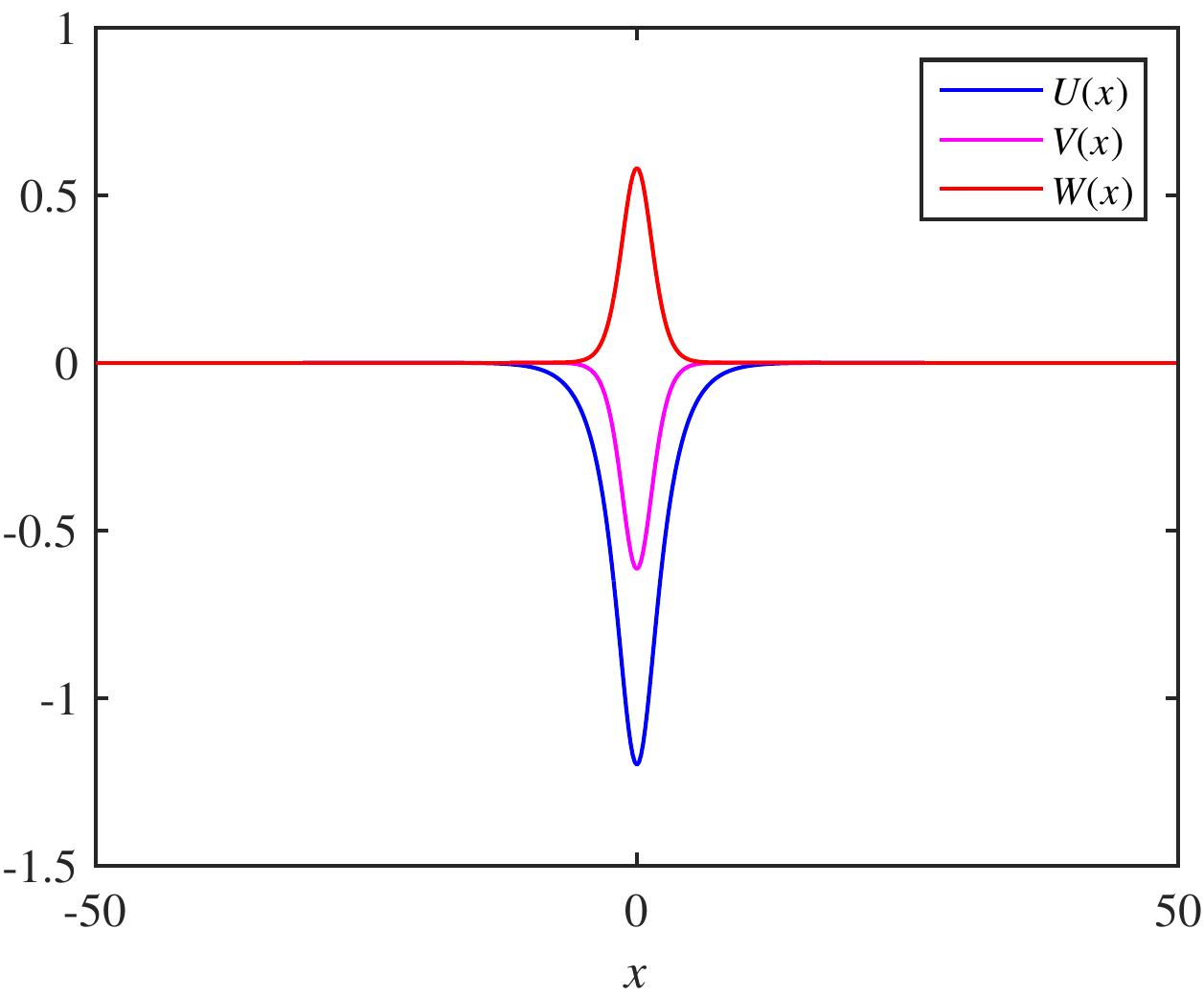}
\caption{(Color online) A generic example of a stable three-component
soliton, with unequal $u$- and $v$-components, found as a solution of Eq. (%
\protect\ref{UVW}) (with $\protect\xi $ replaced by $x$) for parameters $q=-1
$, $b=0.2$, and propagation constants $k_{1}=1/3$ and $k_{2}=2/3$. The same
solution may be taken with opposite signs of $U$ and $V$.}
\label{fig1}
\end{figure}

General three-component soliton solutions to Eq. (\ref{UVW}), without
assuming the symmetry between the two FF components, can be easily
constructed by means of the Newton's method. Numerical computations were
carried out in the domain of $|x|\leq 2^{11}$ with zero boundary conditions,
discretized by $2^{13}$ grid points, achieving a limitation of the residual
relative error $<10^{-10}$. The Newton's method was implemented for fixed $%
k_{1}$ and $k_{2}$, starting from the simplest input which was suggested by
the structure of the exact symmetric solution (\ref{exact}), \textit{viz}., $%
u_{\mathrm{in}}=v_{\mathrm{in}}=-w_{\mathrm{in}}=u_{0}\exp (-x^{2})$, with
some constant $u_{0}$. A typical example of the so found asymmetric solitons
is displayed in Fig. \ref{fig1}.

\begin{figure}[tbp]
\centering\subfigure[]{\includegraphics[width=3in]{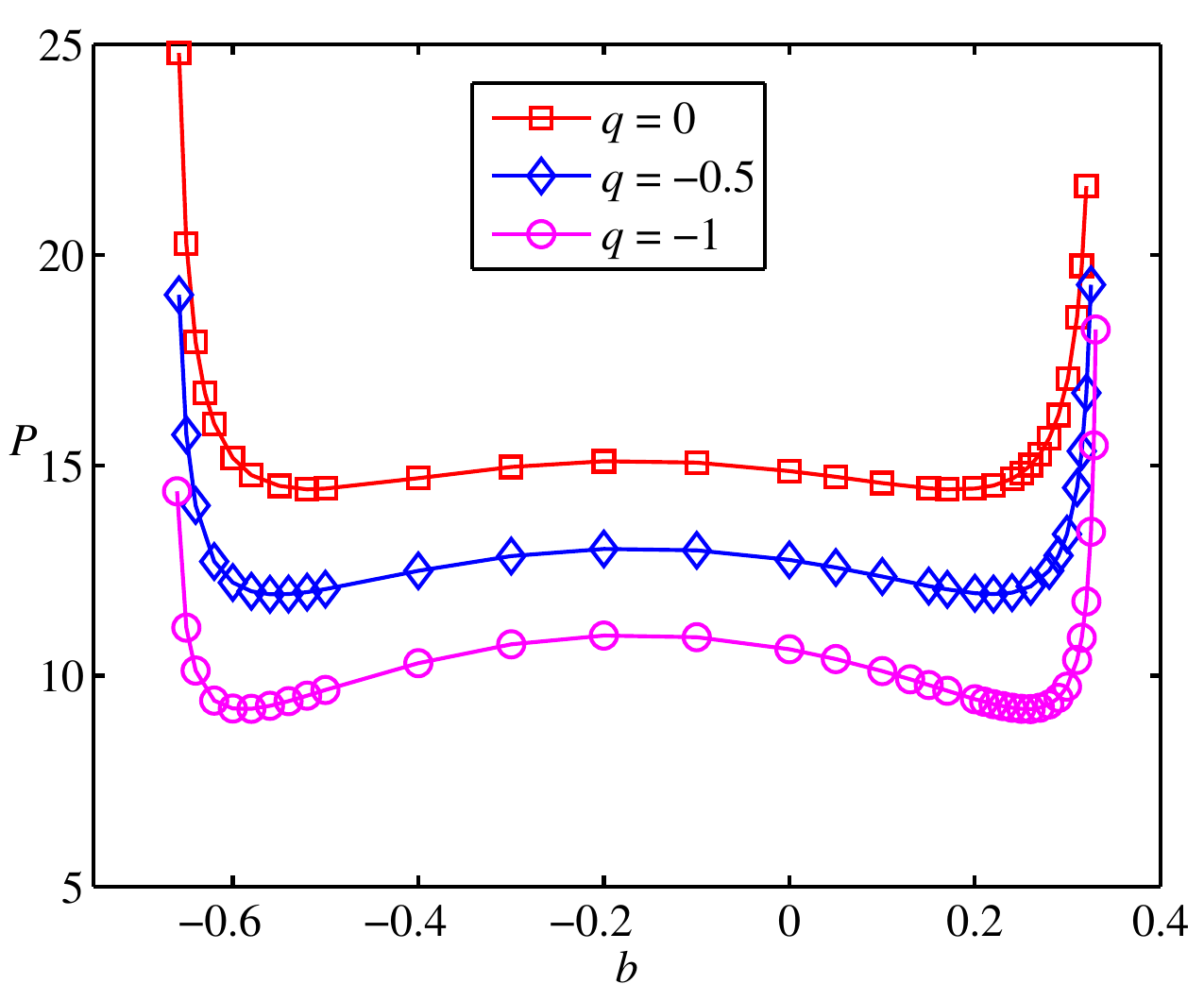}}%
\subfigure[]{\includegraphics[width=3in]{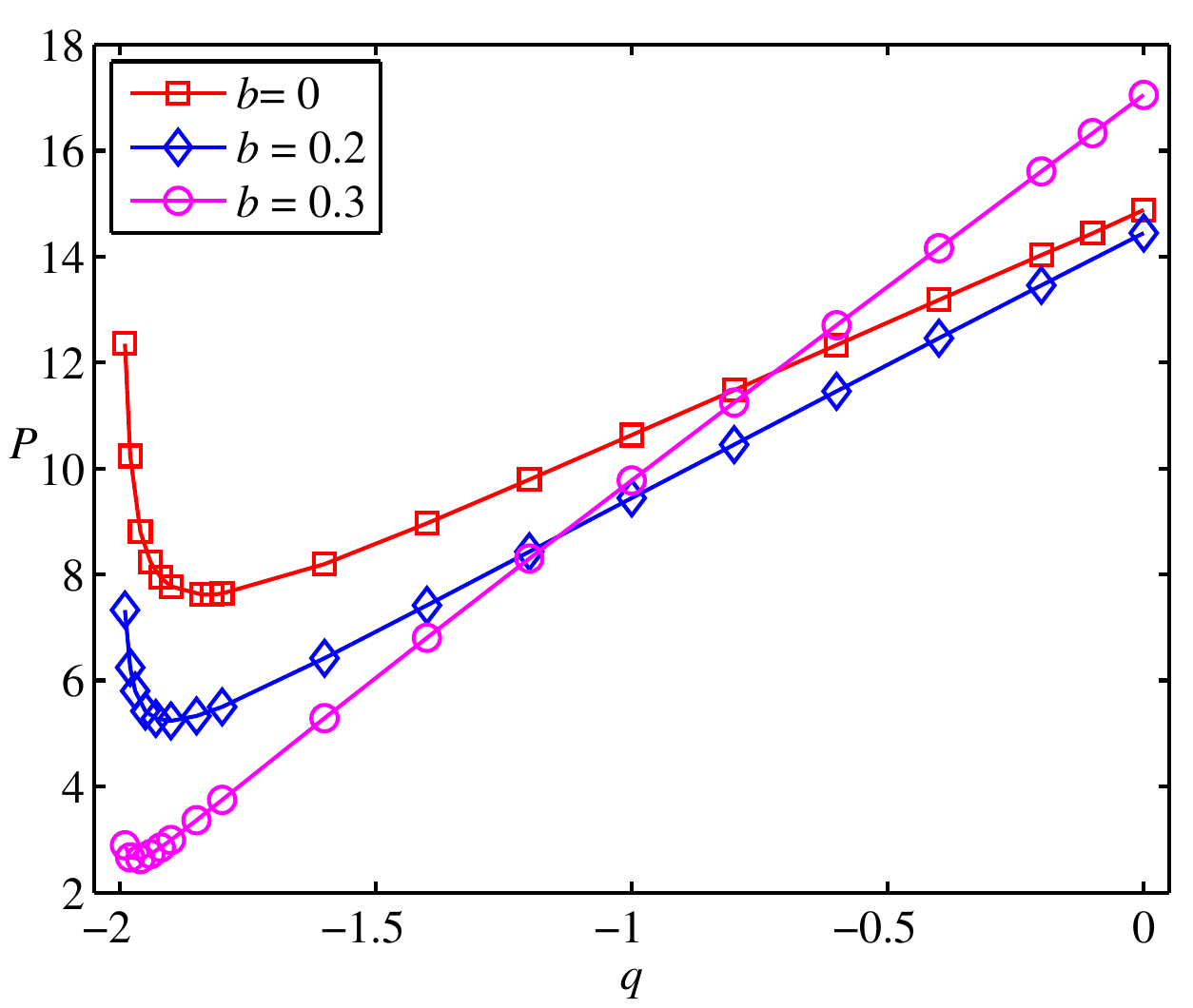}}
\caption{(Color online) (a) The total power $P$ of the three-wave solitons
versus the birefringence coefficient, $b$, for three different values of
mismatch $q$. (b) $P$ versus $q$ for three different values of $b$. In both
panels, the propagation constants are fixed, $k_{1}=1/3$ and $k_{2}=2/3$,
see Eq. (\protect\ref{UVW}).}
\label{fig2}
\end{figure}

Families of generic asymmetric solitons are represented by dependences $P(b)$
for several values of $q$, and $P(q)$ for several values of $b$, which are
displayed in Fig. \ref{fig2} for fixed values of both propagation constants,
$k_{1}=1/3$ and $k_{2}=2/3$. In particular, conditions (\ref{<<>}) determine
the left and right edges of the existence region of the soliton families in
Fig. \ref{fig2}(a), which are, respectively, $b=-k_{2}=-2/3$ and $b=k_{1}=1/3
$, as well as the left edge in Fig. \ref{fig2}(b), $q=-2\left(
k_{1}+k_{2}\right) =-1$. Further, the nearly linear form of the right branch
of the $P(q)$ dependence in Fig. \ref{fig2}(b) can be easily explained by
the cascading approximation, which assumes that the diffraction term (second
derivative) may be neglected in the equation for $W$ \cite{rev1}-\cite{rev3}%
. Indeed, eliminating the SH field under this assumption,
\begin{equation}
W\approx UV/Q,~Q\equiv 2\left( k_{1}+k_{2}\right) +q,  \label{Q}
\end{equation}%
one arrives at a system of equations for $U$ and $V$,%
\begin{gather}
\frac{1}{2}\frac{d^{2}U}{d\xi ^{2}}+\frac{1}{Q}V^{2}U=\left( k_{1}-b\right)
U,  \notag \\
\frac{1}{2}\frac{d^{2}V}{d\xi }+\frac{1}{Q}U^{2}V=\left( k_{2}+b\right) V.
\label{cascading}
\end{gather}%
It is obvious that the total norm of solutions of system (\ref{cascading})\
scales $\sim Q$, i.e., as a linear function of $q$.

As shown below, collisions between TAWs often generate strongly asymmetric
solitons, with one FF component being much taller and broader than the
other, and a still smaller SH component. It is easy to obtain an approximate
form of the strongly asymmetric solitons from Eq. (\ref{cascading}). If, for
instance, the $U$ and $V$ components are the large and small ones, the
effective small-amplitude narrow potential in the first equation in system (%
\ref{cascading}), $-Q^{-1}V^{2}(x)$, may be approximated by a
delta-function. This approximation yields the following result:%
\begin{equation}
U(x)\approx \sqrt{Q\left( k_{2}+b\right) }\exp \left( -\sqrt{2\left(
k_{1}-b\right) }|x|\right) ,~V^{2}(x)\approx Q\sqrt{2\left( k_{1}-b\right) }%
\delta (x),  \label{UV}
\end{equation}%
with the respective powers
\begin{equation}
P_{u}=\frac{Q\left( k_{2}+b\right) }{\sqrt{2\left( k_{1}-b\right) }},~P_{v}=Q%
\sqrt{2\left( k_{1}-b\right) },  \label{PP}
\end{equation}%
while the power of the SH component is negligible in this approximation. The
underlying condition of the dominance of the $U$ component implies $%
P_{v}/P_{u}\ll 1$, i.e., $k_{1}-b\ll k_{2}+b$. The soliton with large $V$
and small $U$ components can be obtained from here by substitution $%
k_{2}\rightleftarrows k_{1},b\rightarrow -b$. A detailed consideration
demonstrates that Eq. (\ref{PP}) and its counterpart, corresponding to the
swap between $U$ and $V$, correctly approximate the right and left peaks of
the dependences displayed in Fig. \ref{fig2}(a).

Stability of the three-component solitons was tested by systematic
simulations of their perturbed evolution, using the split-step
fast-Fourier-transform algorithm. The simulations were typically performed
in the propagation interval $0\leq z\leq 500$, which roughly corresponds to $%
20$ diffraction lengths of the typical soliton displayed in Fig. \ref{fig1}.
In the course of the simulations, the conservation of the total power (\ref%
{P1D}) and momentum \ref{M1D} holds with relative accuracy no worse than $%
10^{-5}$. As a result, it has been concluded that the entire family of exact
solitons (\ref{k12}), (\ref{exact}) is stable, as well as all the asymmetric
solitons found in the numerical form.

\begin{figure}[tbp]
\centering\subfigure[]{\includegraphics[width=3in]{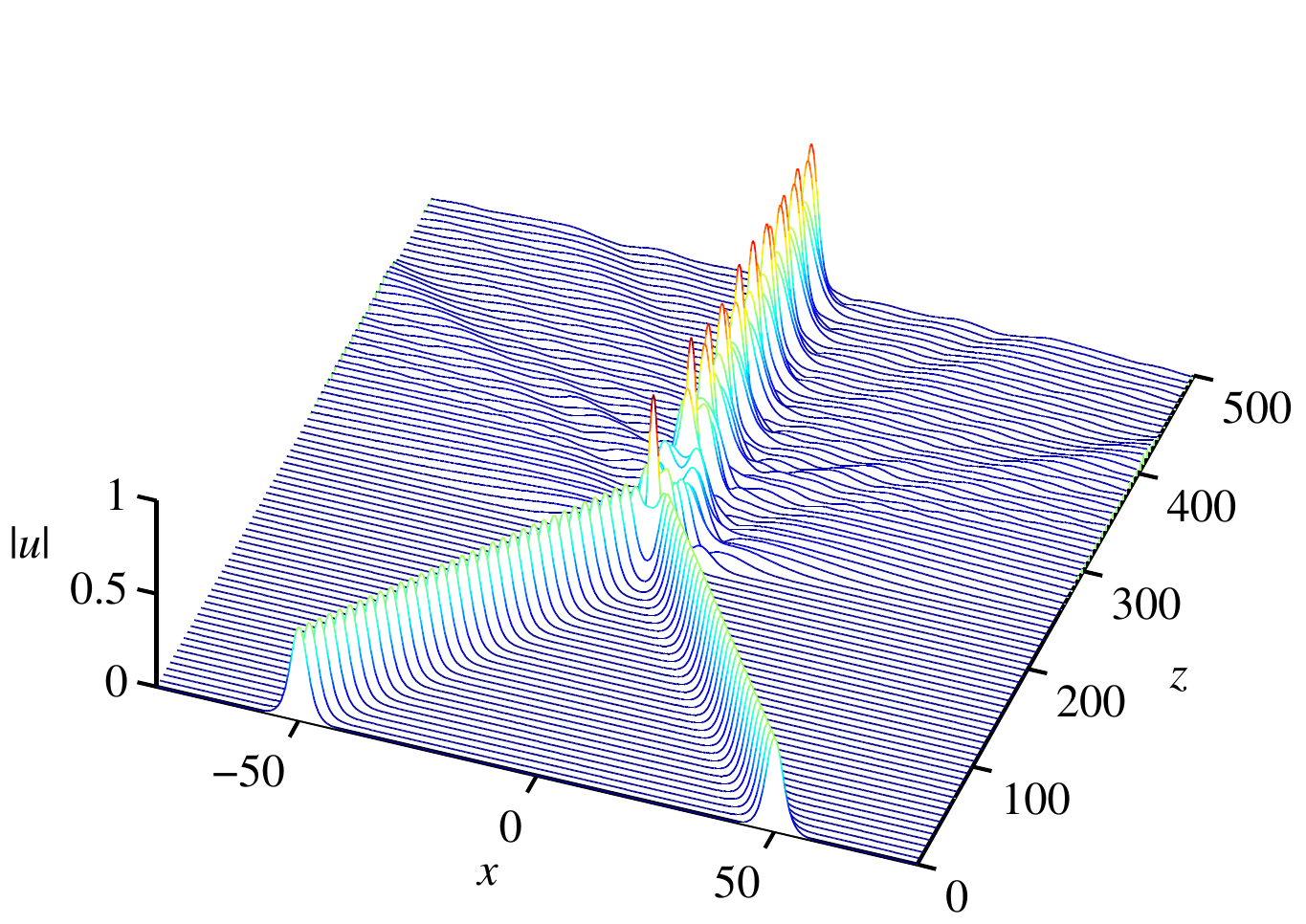}}%
\subfigure[]{\includegraphics[width=3in]{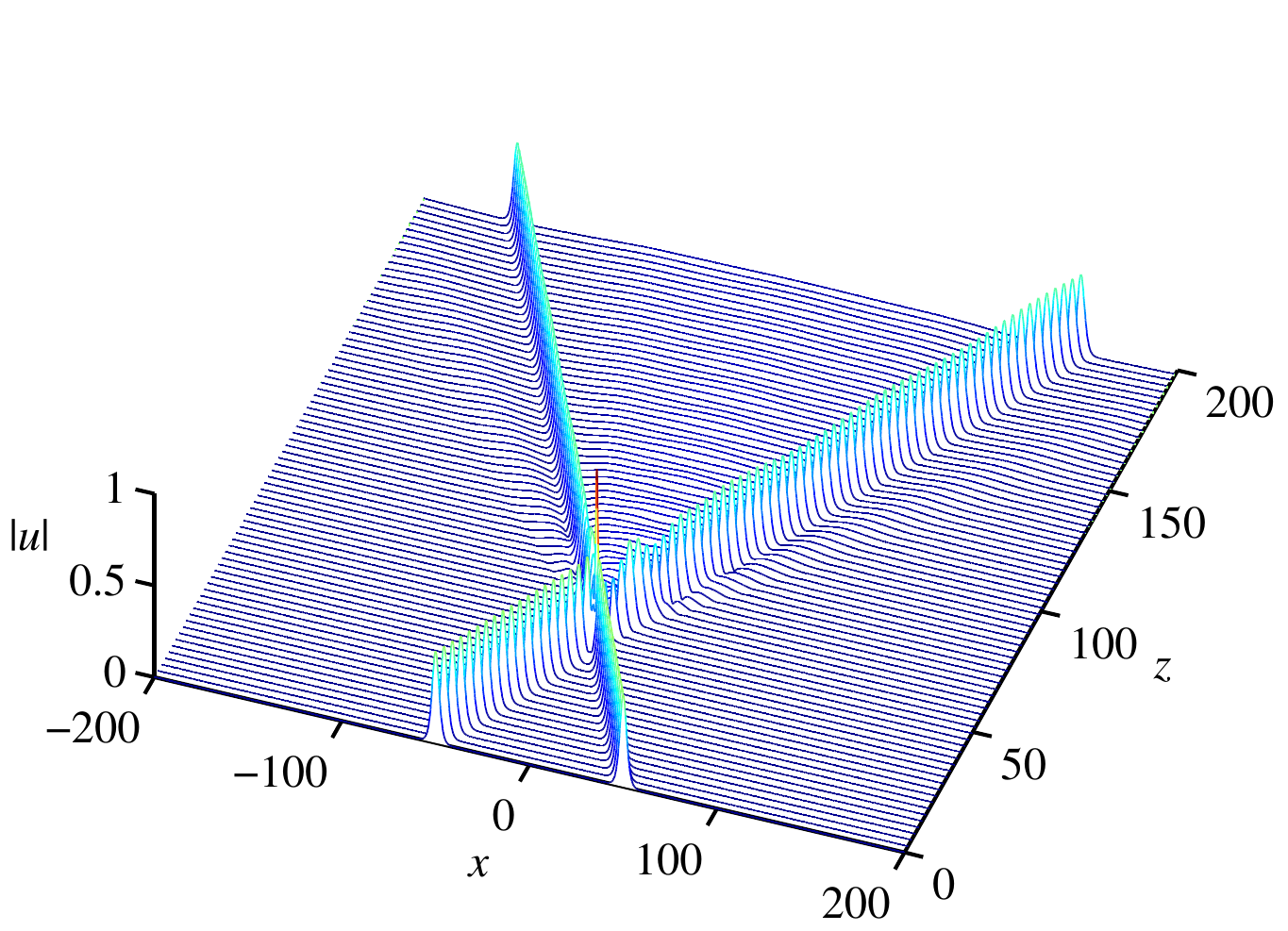}}
\caption{(Color online) (a) Merger of three-wave solitons, given by exact
solution (\protect\ref{k12})-(\protect\ref{exact}), which collide with
velocities $c=\pm 0.23$. (b) A quasi-elastic collision of solitons moving
with velocities $c=\pm 1$. The initial distance between the solitons is $%
\Delta x=100$; other parameters are $b=0$ and $q=-1$.}
\label{fig3}
\end{figure}

Before proceeding to the consideration of collisions of solitons with TAWs,
and collisions of TAWs between themselves, it is relevant to briefly
consider collisions between moving solitons. The simulations demonstrate a
situation typical for interactions of solitons in nonintegrable systems \cite%
{RMP}: collisions between relatively slow solitons lead to their merger into
a single excited localized state, which is accompanied by the emission of
small-amplitude dispersive waves, while fast solitons interact
quasi-elastically, passing through each other, as shown, in terms of the
evolution of field $|u\left( x,z\right) |$, in Figs. \ref{fig3}(a) and (b),
respectively (the evolution of the other fields, $v$ and $w$, shows a
similar picture). For values of parameters fixed in this figure, $b=0$ and $%
q=-1$, the critical velocity which separates the inelastic and elastic
collisions of the solitons moving with velocities $\pm c$ is $c_{\mathrm{cr}}
$ $\approx 0.245$.

\section{Collisions between counter-propagating Airy waves}

\begin{figure}[tbp]
\centering\subfigure[]{\includegraphics[width=3in]{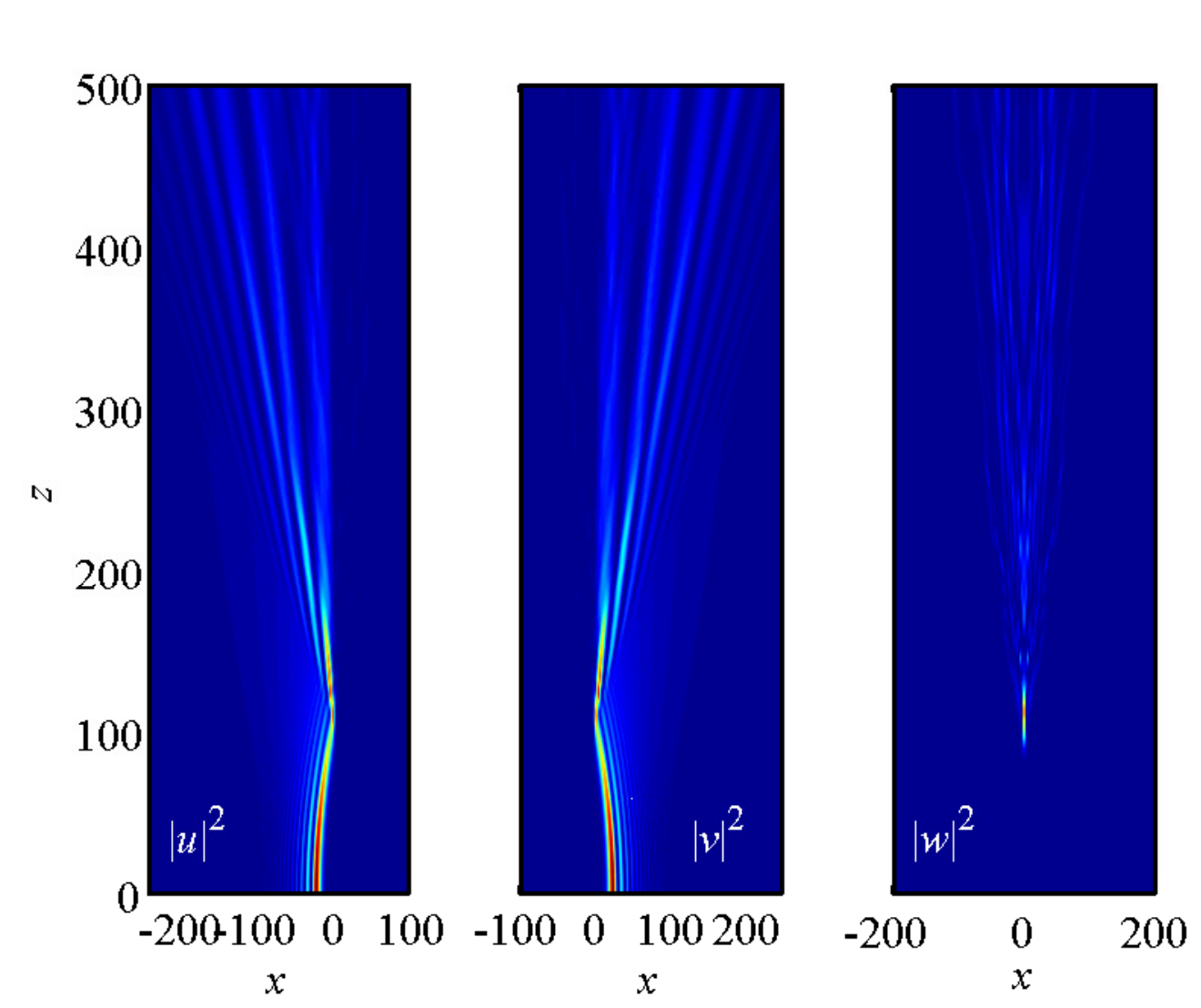}}%
\subfigure[]{\includegraphics[width=3in]{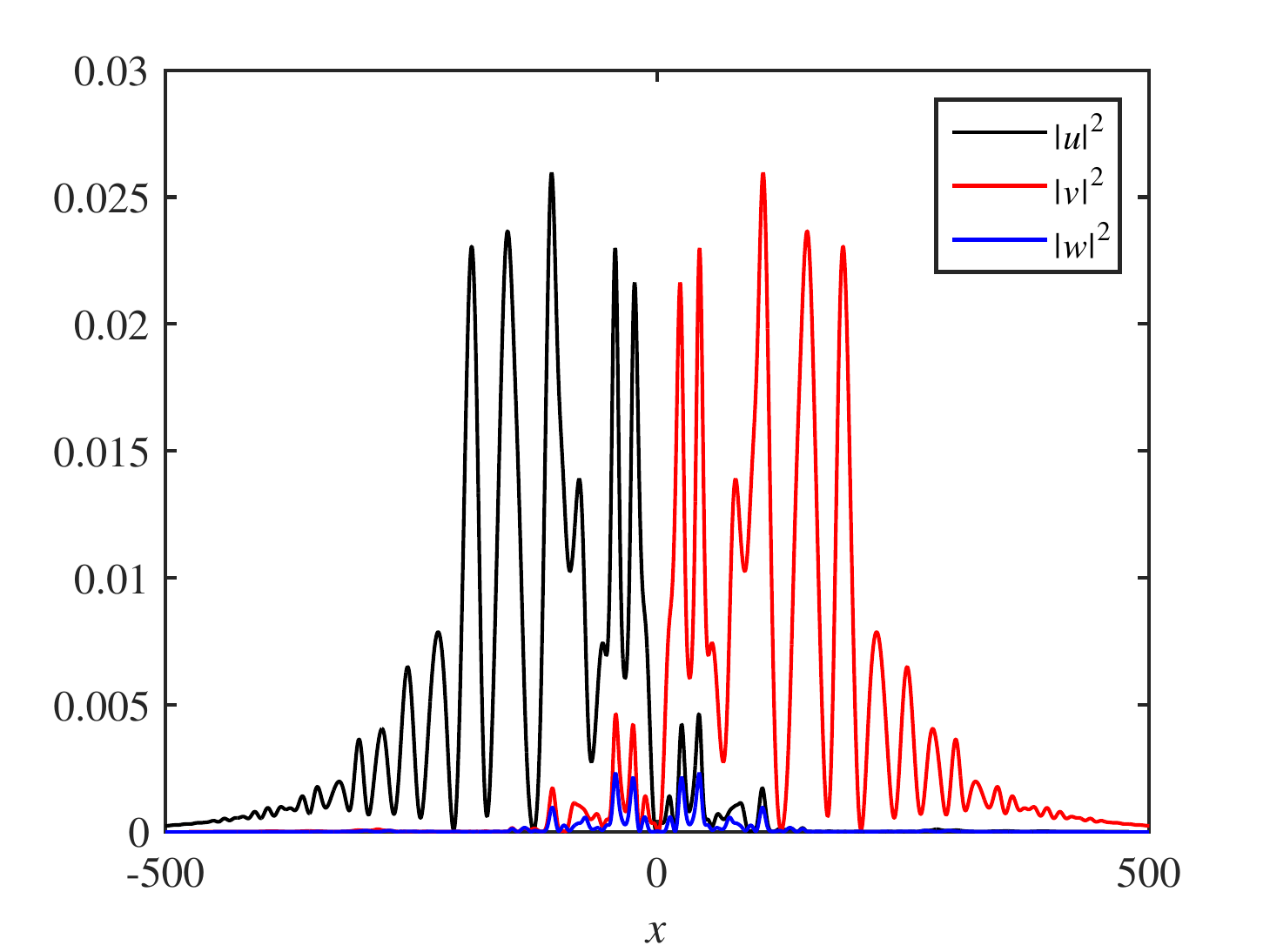}}
\caption{(Color online) (a) The collision between two Airy waves, (\protect
\ref{1D-AW}) and (\protect\ref{tilde}), launched with parameters $\protect%
\alpha =\pm 0.2,\aleph =0.1,u_{0}=1$, is displayed by means of the evolution
of local powers $\left\vert u\left( x,z\right) \right\vert ^{2}$, $%
\left\vert v\left( x,z\right) \right\vert ^{2}$, and $\left\vert w\left(
x,z\right) \right\vert ^{2}$. (b) The final power profiles of fields $u$, $v$
and $w$ at $z=500$, which demonstrate the transformation of the colliding
waves into a complex multi-peak pattern.}
\label{fig4}
\end{figure}

\begin{figure}[tbp]
\centering\subfigure[]{\includegraphics[width=3in]{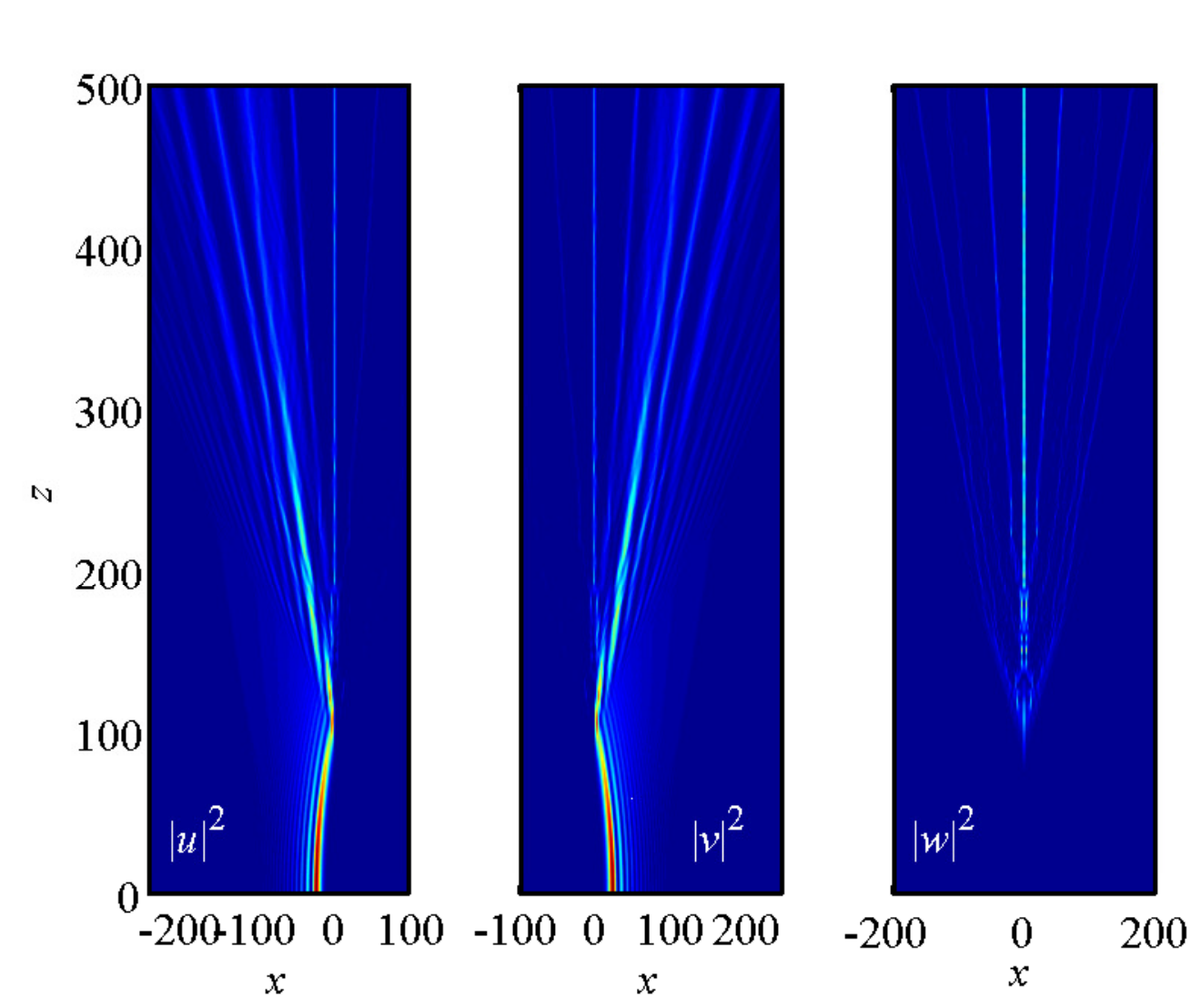}}%
\subfigure[]{\includegraphics[width=3in]{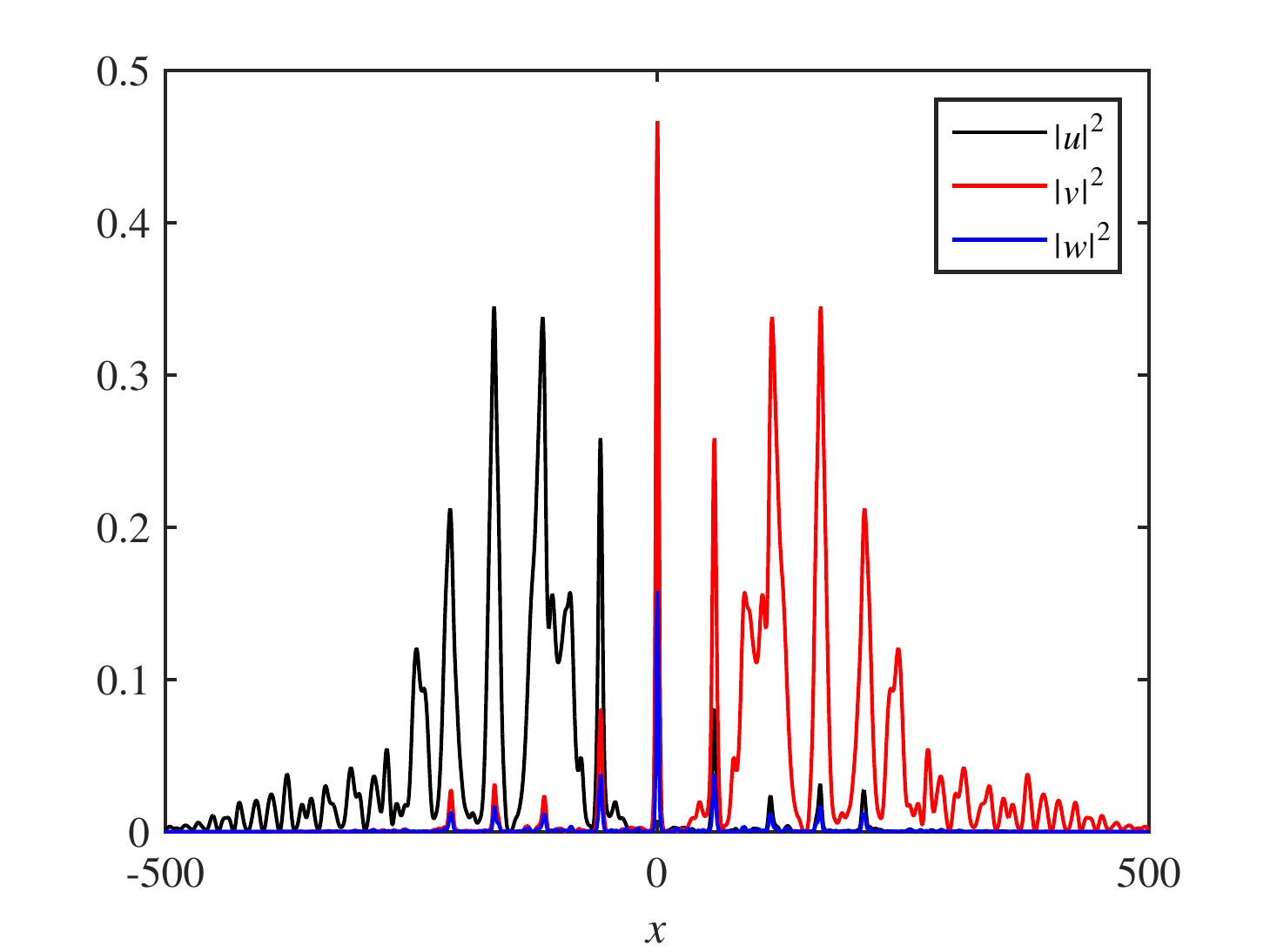}}
\caption{(Color online) The same as in Fig. \protect\ref{fig4}, but for $%
u_{0}=3$. In these plots and similar ones, displayed below for $u_{0}=5$ and
$6$ in Figs. \protect\ref{fig6}(b) and \protect\ref{fig7}(b), isolated peaks
can be identified as three-wave solitons, see the text.}
\label{fig5}
\end{figure}

\begin{figure}[tbp]
\centering\subfigure[]{\includegraphics[width=3in]{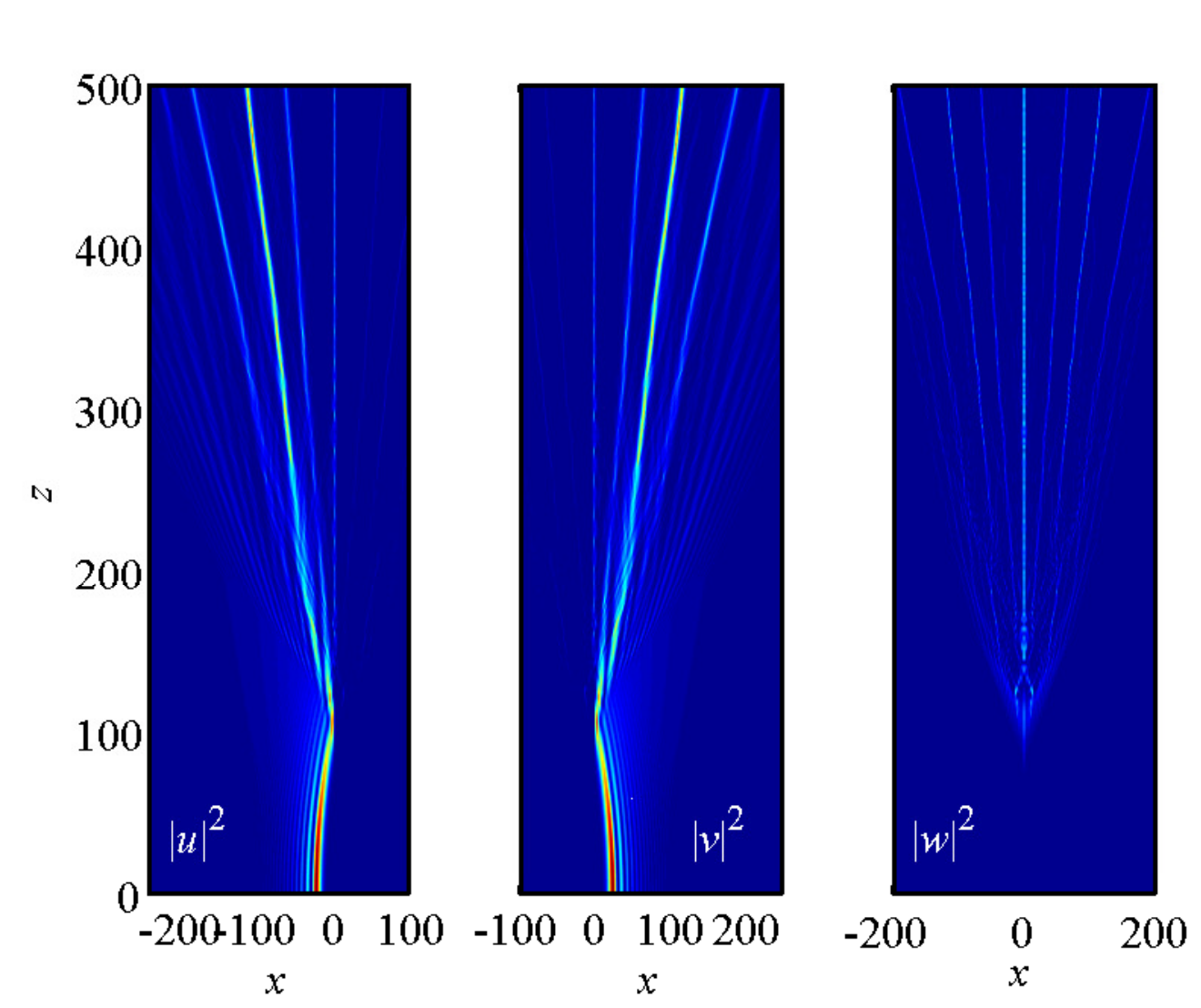}}%
\subfigure[]{\includegraphics[width=3in]{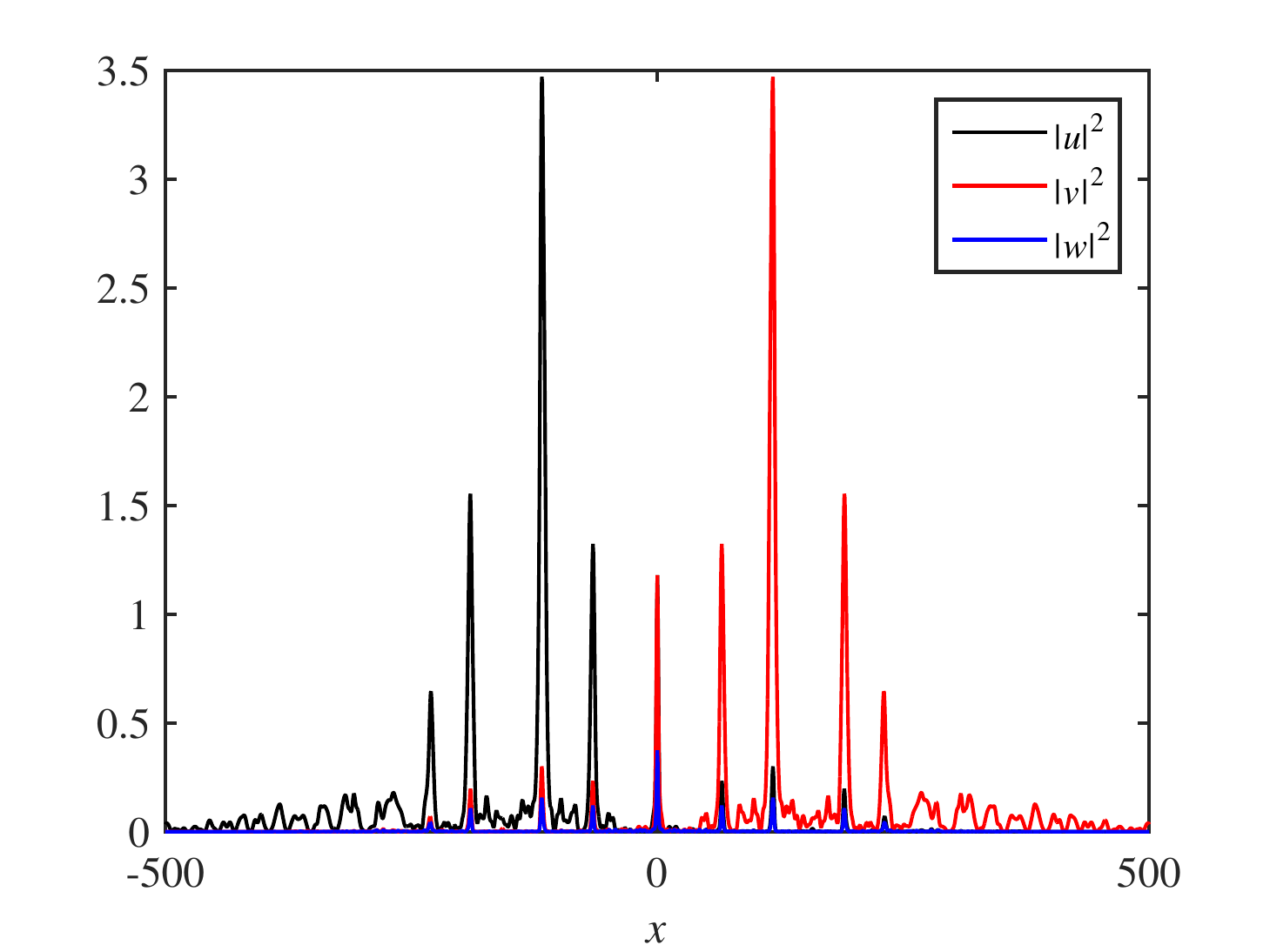}}
\caption{(Color online) The same as in Figs. \protect\ref{fig4} and \protect
\ref{fig5}, but for $u_{0}=5$.}
\label{fig6}
\end{figure}

\begin{figure}[tbp]
\centering\subfigure[]{\includegraphics[width=3in]{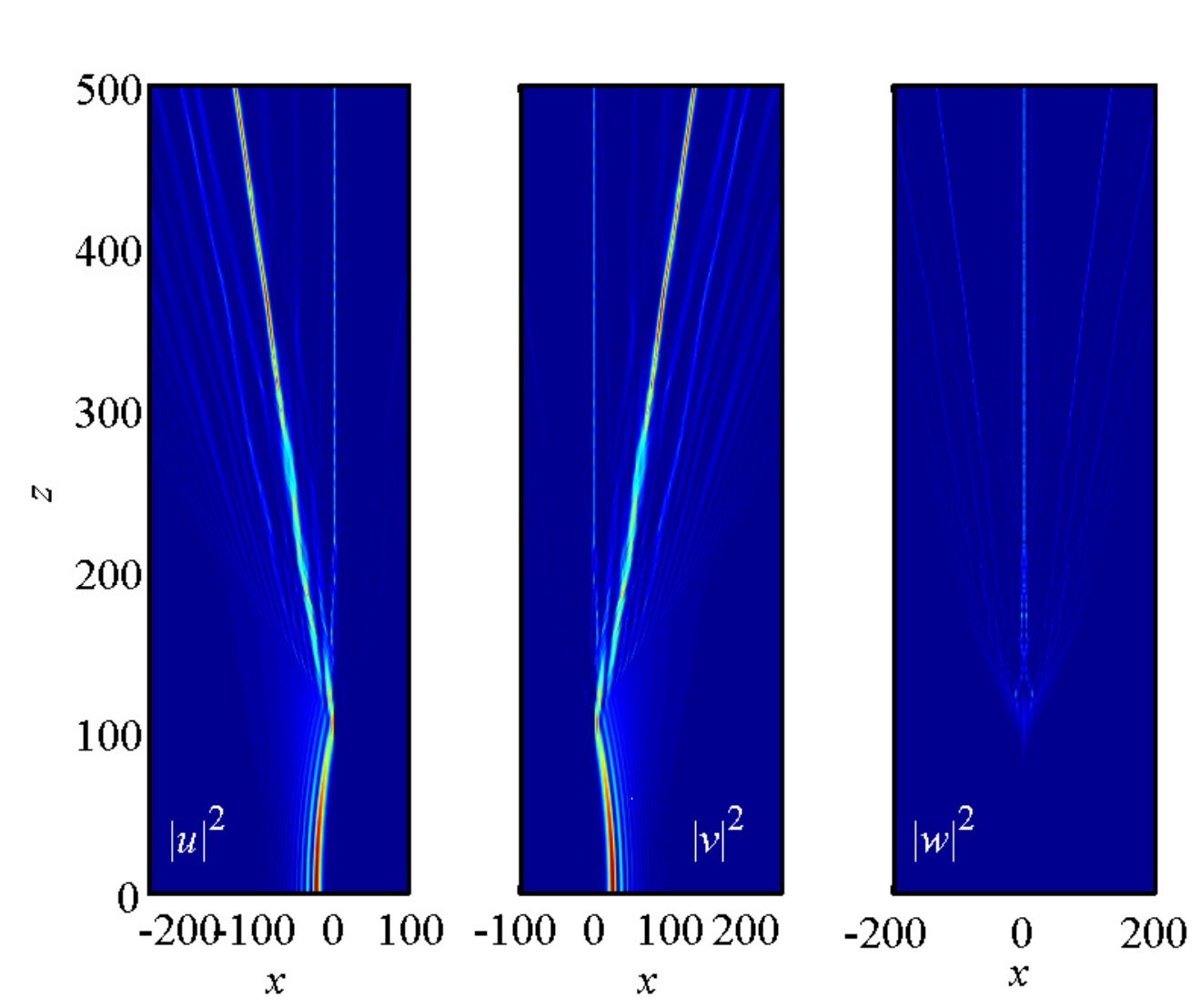}}%
\subfigure[]{\includegraphics[width=3in]{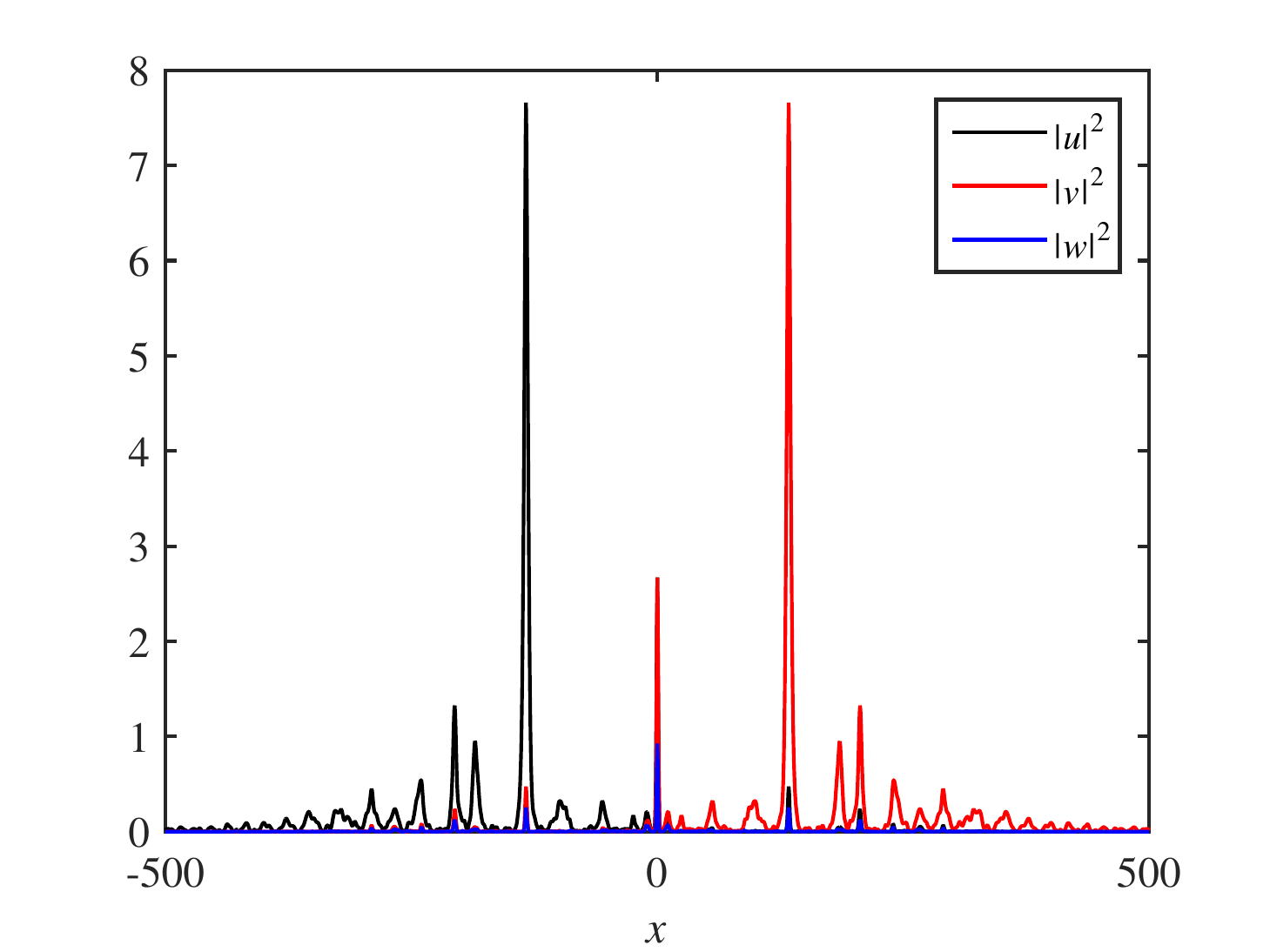}}
\caption{(Color online) The same as in Figs. \protect\ref{fig4}-\protect\ref%
{fig6}, but for $u_{0}=6$.}
\label{fig7}
\end{figure}

As mentioned above, the three-wave system makes it possible to consider
collisions between undistorted TAWs, such as one given by Eq. (\ref{1D-AW})
and the counterpropagating one (\ref{tilde}). The result of the collision
strongly depends on the strength of the nonlinearity, which is determined by
amplitude $u_{0}$ in Eq. (\ref{1D-AW}). This is illustrated, in a generic
form, in Figs. \ref{fig4}-\ref{fig7} by a set of results generated for the
pair of the TAWs with $\alpha =\pm 0.2,\aleph =0.1$, launched with initial
separation $X=40$ between their centers, while parameters in Eq. (\ref%
{type-II}) are taken as $b=$ $q=0$. In principle, the outcome of the
collision may also depend on a phase shift between the two mutually
symmetric TAWs \cite{Belic'}; however, the simulations demonstrate that this
effect is inconspicuous, due to the fact that the variation of the phase
across the TAWs given by Eqs. (\ref{1D-AW}) and (\ref{tilde}) becomes larger
than the phase shift between them.

First, in the quasi-linear regime, which corresponds to a relatively small
amplitude, $u_{0}=1$, Fig. \ref{fig4} demonstrates transformation of the
interacting Airy waves into a complex multi-peak pattern, in which
individual peaks may be roughly traced backs to lobes of the colliding TAWs.
The increase of the amplitude from $u_{0}=1$ to $3$ leads to coagulation of
many peaks into fewer taller ones, due to the nonlinear self-attraction, as
seen in Fig. \ref{fig5}. The trend continues with further increase of $u_{0}$%
, producing the cluster of peaks displayed in Fig. \ref{fig6} for $u_{0}=5$.
In the latter case, detailed consideration definitely confirms that each
isolated tall peak may be identified as a stable three-wave soliton -- a
quiescent symmetric one at the center, and eight strongly asymmetric moving
solitons, which may be approximated by Eqs. (\ref{UV}) and (\ref{PP}), with
the Galilean boost ($\sim c$) applied to them, as per Eq. (\ref{sol}).

Eventually, under the action of sufficiently strong nonlinearity, the
collision of two TAWs leads to the formation of precisely three solitons:
the central symmetric one, and a pair of strongly asymmetric left- and
right-traveling solitons. This ultimate outcome of the collisions is
achieved by increasing the input amplitudes from $u_{0}=5$ in Fig. \ref{fig6}%
, where nine solitons are still clearly identified in the output, to $u_{0}=6
$ (i.e., increasing the total power by $44\%$), as shown in Fig. \ref{fig7}.

It is relevant to note that, in the case of the sufficiently strong
nonlinearity, the origin of individual solitons in the output of the
collision cannot be traced back to main lobes of the initial TAW profiles.
In particular, additional simulations (not shown here in detail) demonstrate
that, in the moderately nonlinear situation, the number of generated
solitons is essentially affected by varying $\alpha $, i.e., the intrinsic
scale of the TAW in Eq. (\ref{z=0}), but not by changing parameter $\aleph $%
, which determines the number of lobes in the waveform (\ref{z=0}).

\section{Airy-soliton collisions}

Because both\ the single-component TAWs and three-component solitons are
eigenmodes of the three-wave system, it is natural to study interactions
(collisions) between them. To this end, we took the TAW in the form of Eq. (%
\ref{z=0}) with
\begin{equation}
u_{0}=1,\alpha =\aleph =0.1,  \label{0.1}
\end{equation}%
the respective power being $P_{\mathrm{TAW}}\approx 6.29$, as per Eq. (\ref%
{P1DAW}). The main lobe of this Airy wave has width $L_{\mathrm{FWHM}%
}\approx 17.1$. Systematic simulations demonstrate that this choice of the
parameters adequately represents the generic situation.

\begin{figure}[tbp]
\centering\subfigure[]{\includegraphics[width=3in]{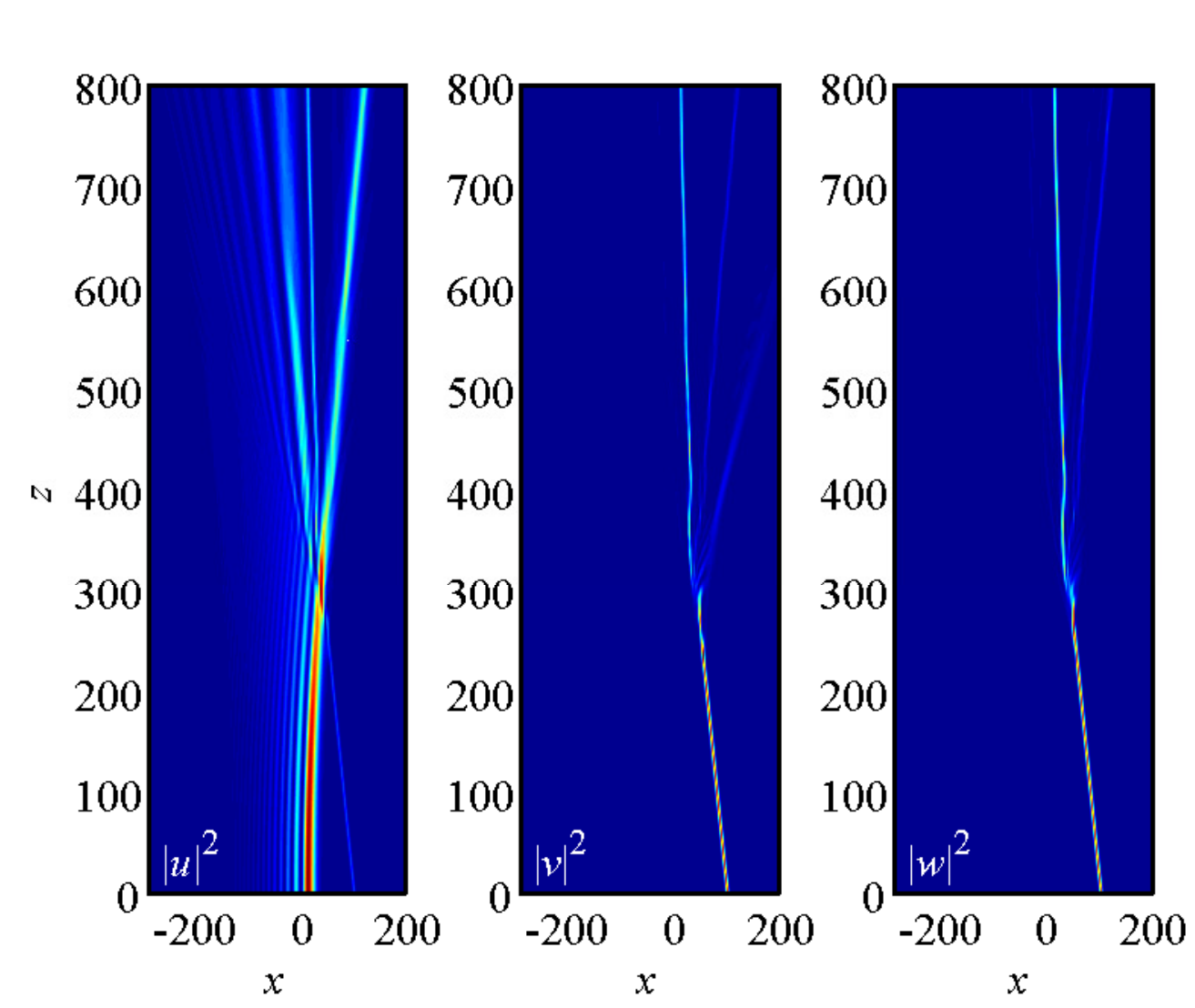}}%
\subfigure[]{\includegraphics[width=3in]{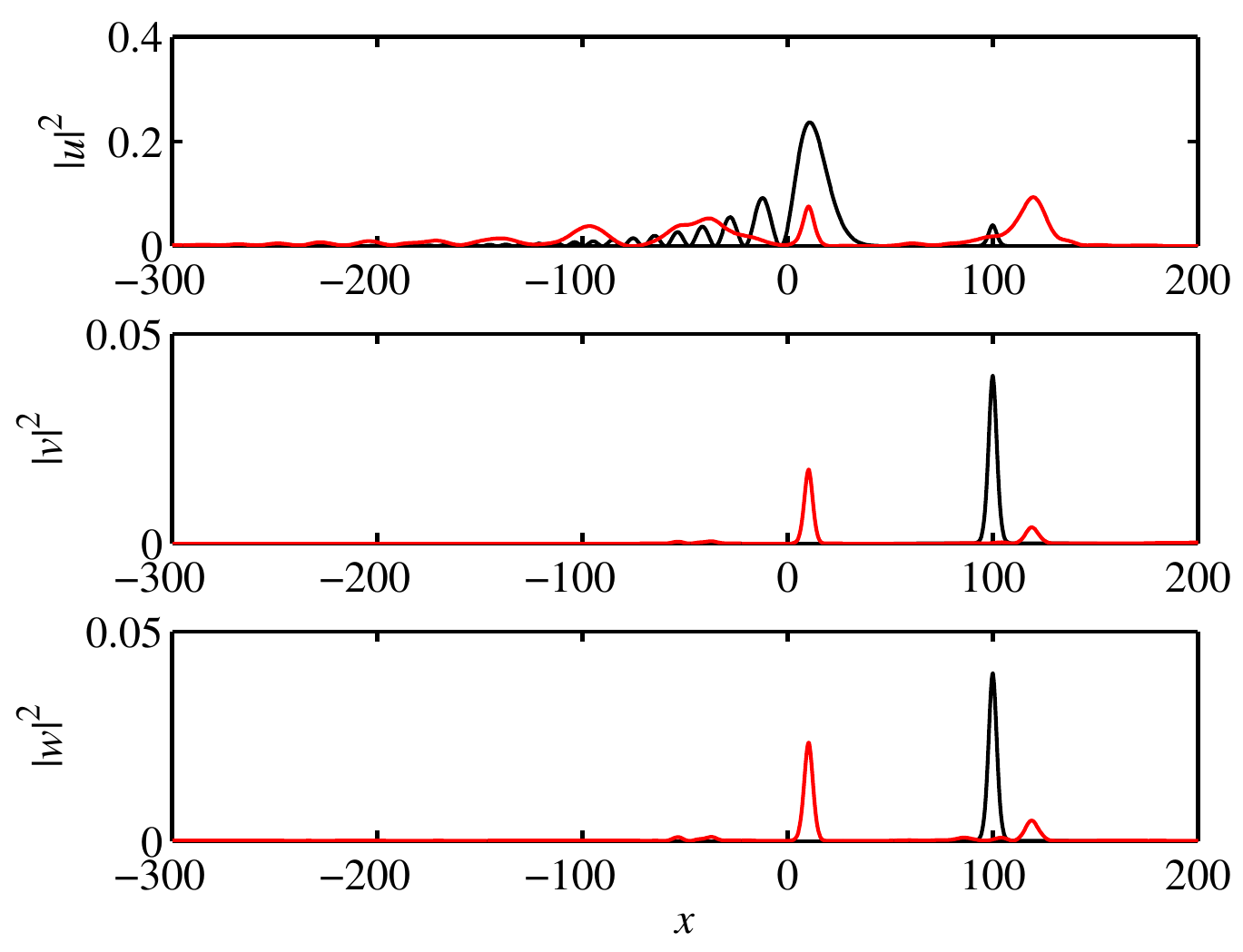}}
\caption{(Color online) (a) The collision between the Airy wave with
parameters (\protect\ref{0.1}) and the symmetric soliton (\protect\ref{exact}%
), at $q=-0.4$, with velocity $c=-0.2$, is displayed by means of the
evolution of local powers $\left\vert u\left( x,z\right) \right\vert ^{2}$, $%
\left\vert v\left( x,z\right) \right\vert ^{2}$, and $\left\vert w\left(
x,z\right) \right\vert ^{2}$. (b) Black and red profiles display,
respectively, initial and final power profiles of fields $u$, $v$ and $w$ at
$z=500$, which demonstrate essential disturbance of the Airy wave produced
by the collision. Note different vertical scales in the three plots in (b),
and in similar plots displayed in subsequent figures below. }
\label{fig8}
\end{figure}

The simulations demonstrate that the TAW is a much more fragile object than
the soliton, which is not surprising, as the TAW has a complex, hence more
vulnerable, structure, and this mode, unlike the soliton, is not adjusted to
the action of the nonlinearity. The first series of simulations was
performed for collisions of the TAW with symmetric solitons (\ref{exact}),
keeping a fixed initial separation between their centers, $\Delta x=100$,
and a relatively small soliton's velocity, $c=-0.2$, that makes it possible
to observe effects of sufficiently strong interactions (if the velocity is
too large, the soliton quickly passes the TAW, which does not allow the
system to accumulate interaction effects, as shown below). When the
soliton's power $P_{\mathrm{sol}}$, given by Eq. (\ref{Psol}), is very
small, one may expect that it may be absorbed by the TAW. This happens
indeed, for $P_{\mathrm{sol}}\lesssim 0.5$, which corresponds to the
soliton's amplitudes $-q\lesssim 0.2$ in Eq. (\ref{exact}). If $-q$ and,
accordingly, $P_{\mathrm{sol}}$ become somewhat larger, Fig. \ref{fig8}
demonstrates that the soliton with $-q=0.4$, i.e., $P_{\mathrm{sol}}\approx
1.24$ (roughly, $1/5$ of $P_{\mathrm{TAW}}$), is still absorbed by the TAW,
which suffers essential disruption, being on the verge of splitting into
solitons.

\begin{figure}[tbp]
\centering\subfigure[]{\includegraphics[width=3in]{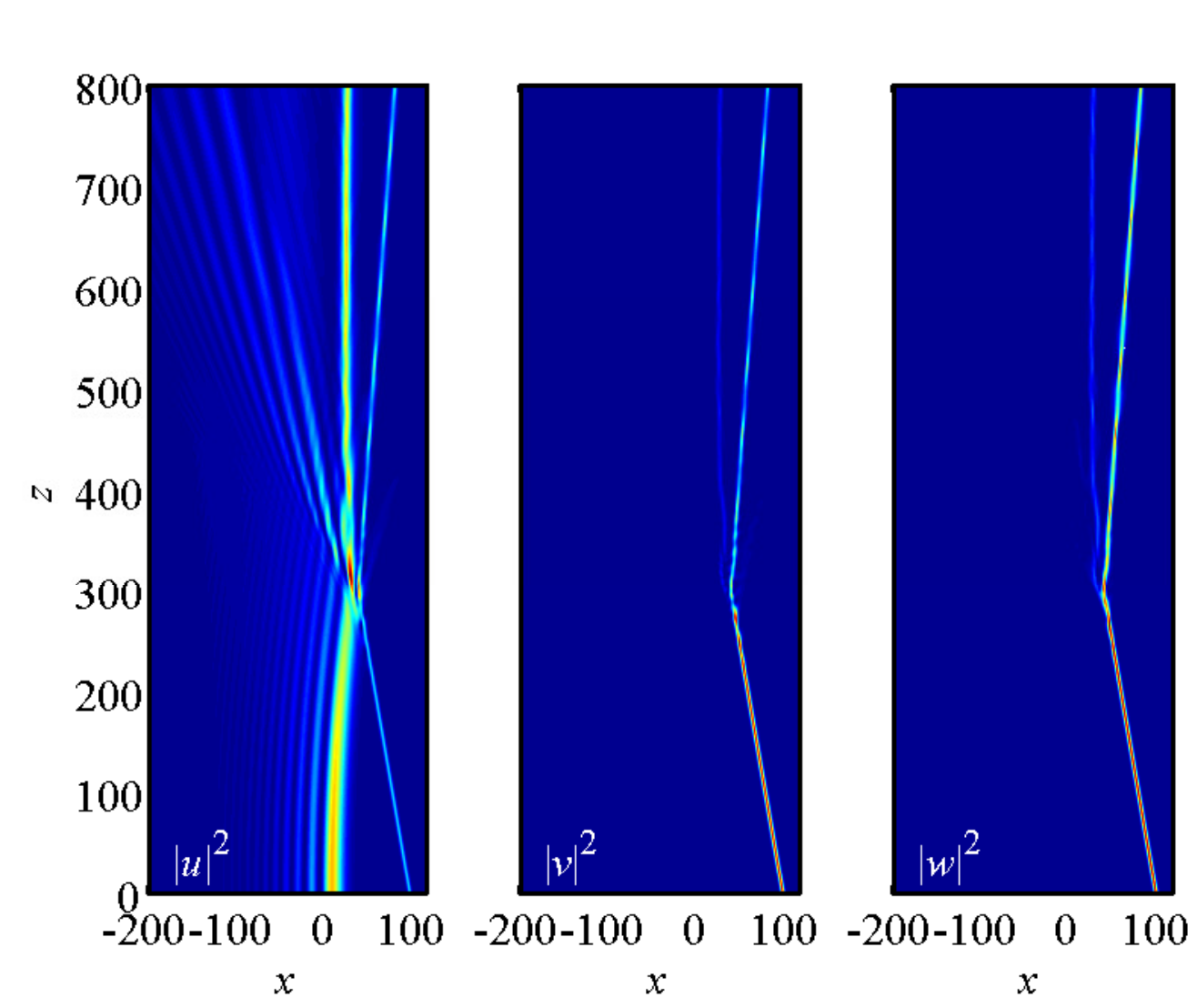}}%
\subfigure[]{\includegraphics[width=3in]{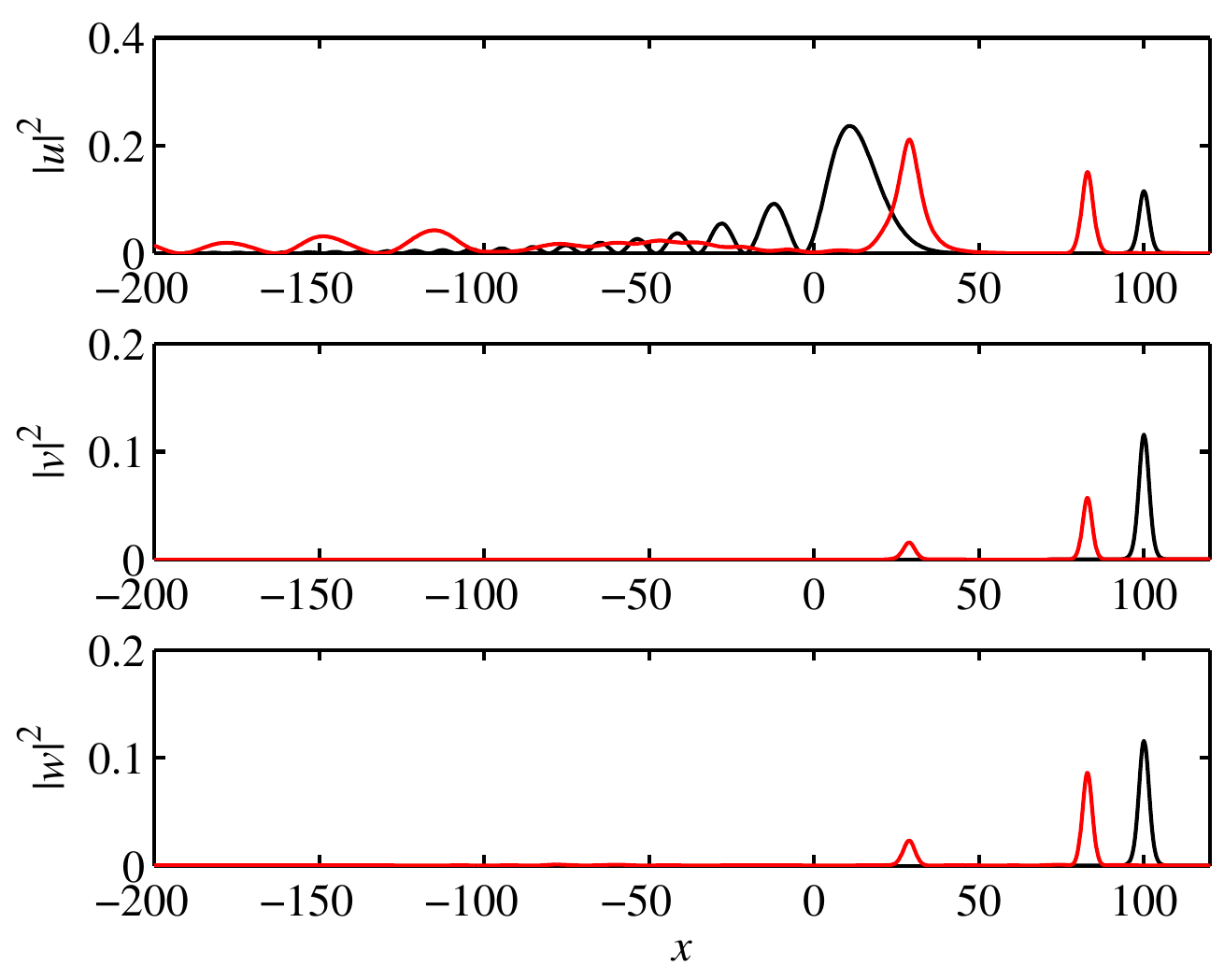}}
\caption{(Color online) The same as in Fig. \protect\ref{fig8}, but for $%
q=-0.68$. In this case, the collision transforms the TAW-soliton pair into a
set of two separating solitons, with a remnant of the Airy wave attached to
one of them.}
\label{fig9}
\end{figure}

\begin{figure}[tbp]
\centering\subfigure[]{\includegraphics[width=3in]{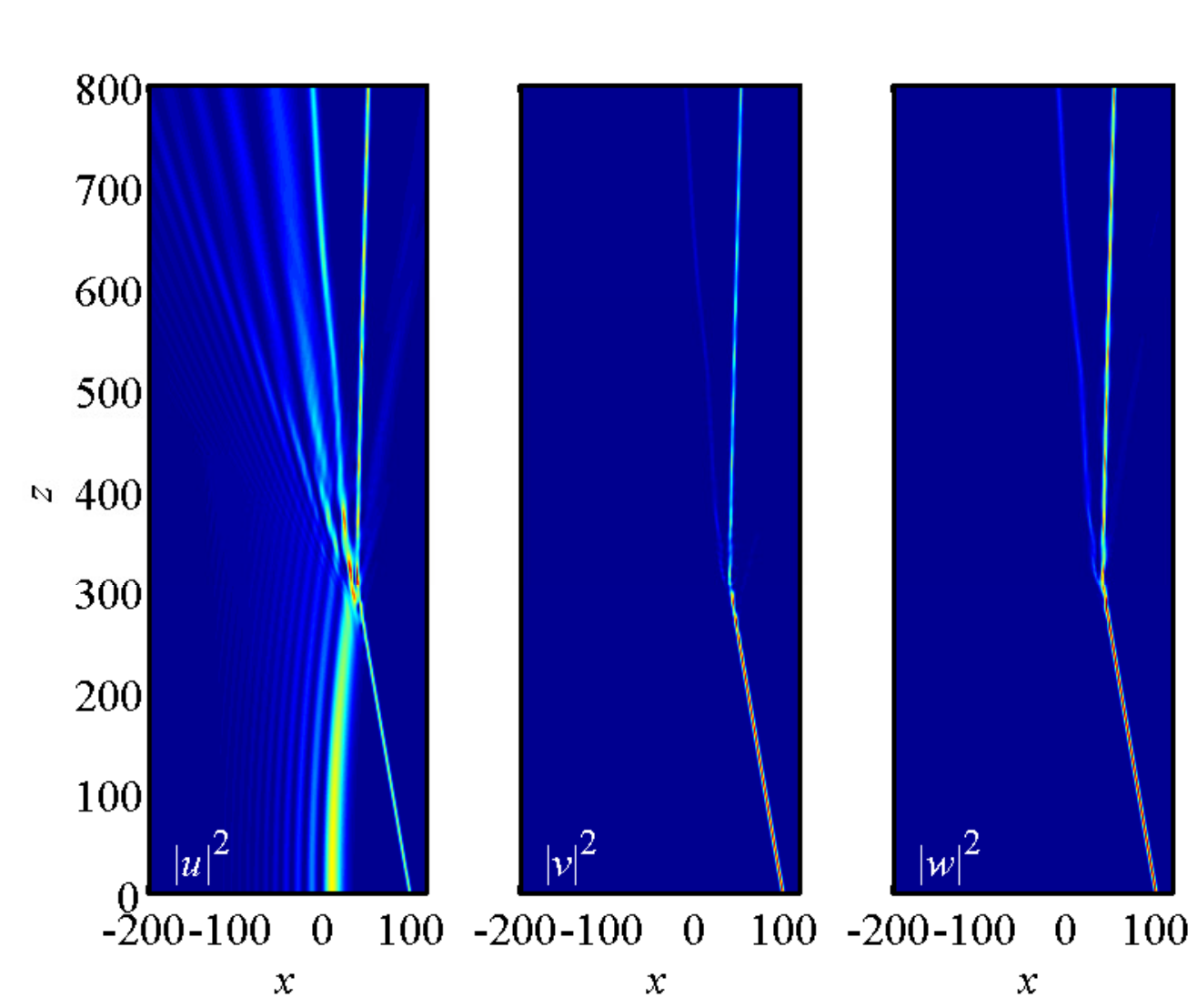}}%
\subfigure[]{\includegraphics[width=3in]{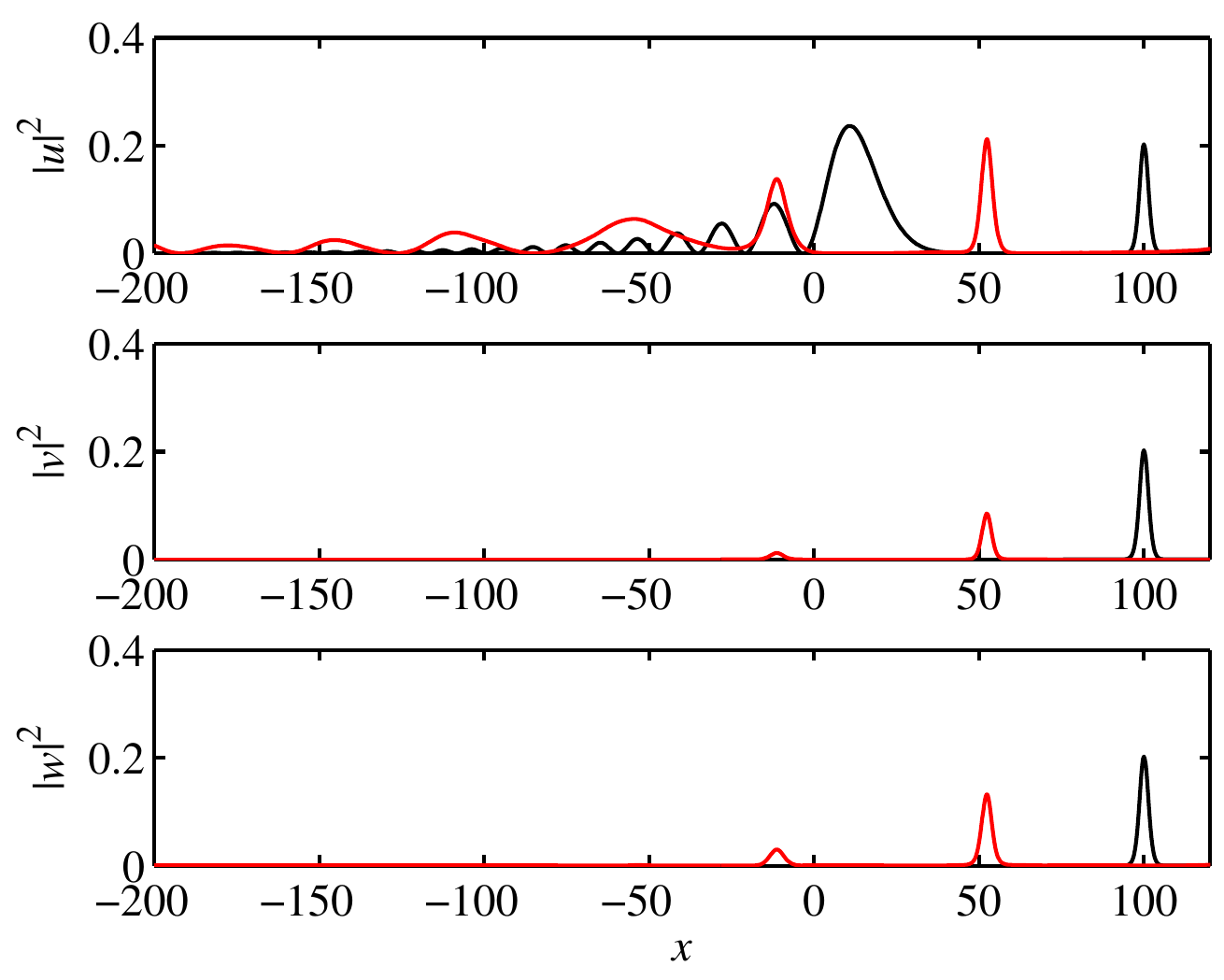}}
\caption{(Color online) The same as in Figs. \protect\ref{fig8} and \protect
\ref{fig9}, but for $q=-0.9$.}
\label{fig10}
\end{figure}

The fragility of the Airy wave becomes apparent at still larger values of $%
P_{\mathrm{sol}}$: the TAW starts to play the role of the reservoir of
power, which is split into fragments by the incident soliton. In particular,
the soliton carrying $P_{\mathrm{sol}}\approx $ \ $\allowbreak 2.74\approx
0.44~P_{\mathrm{TAW}}$, which corresponds to $q=-0.68$, breaks the TAW, as
shown in Fig. \ref{fig9}. In this case, the outcome of the collision is
quite simple: the incident symmetric soliton passes through the TAW,
snatching a large share of the power from the Airy wave, and growing to
become strongly asymmetric. The so amplified asymmetric soliton emerges with
a tail attached on its left side, which may be considered as a remnant of
the Airy wave. The remaining part of the initial TAW's power self-traps into
another asymmetric soliton, which follows the propagation direction of the
original TAW. A qualitatively similar outcome of the collision (which is,
thus, a generic one in certain interval of the values of $P_{\mathrm{sol}}$)
is observed at $P_{\mathrm{sol}}=4.18\approx \allowbreak 0.66~P_{\mathrm{TAW}%
}$, which corresponds to $q=-0.9$, as seen in Fig. \ref{fig10}.

\begin{figure}[tbp]
\centering\subfigure[]{\includegraphics[width=3in]{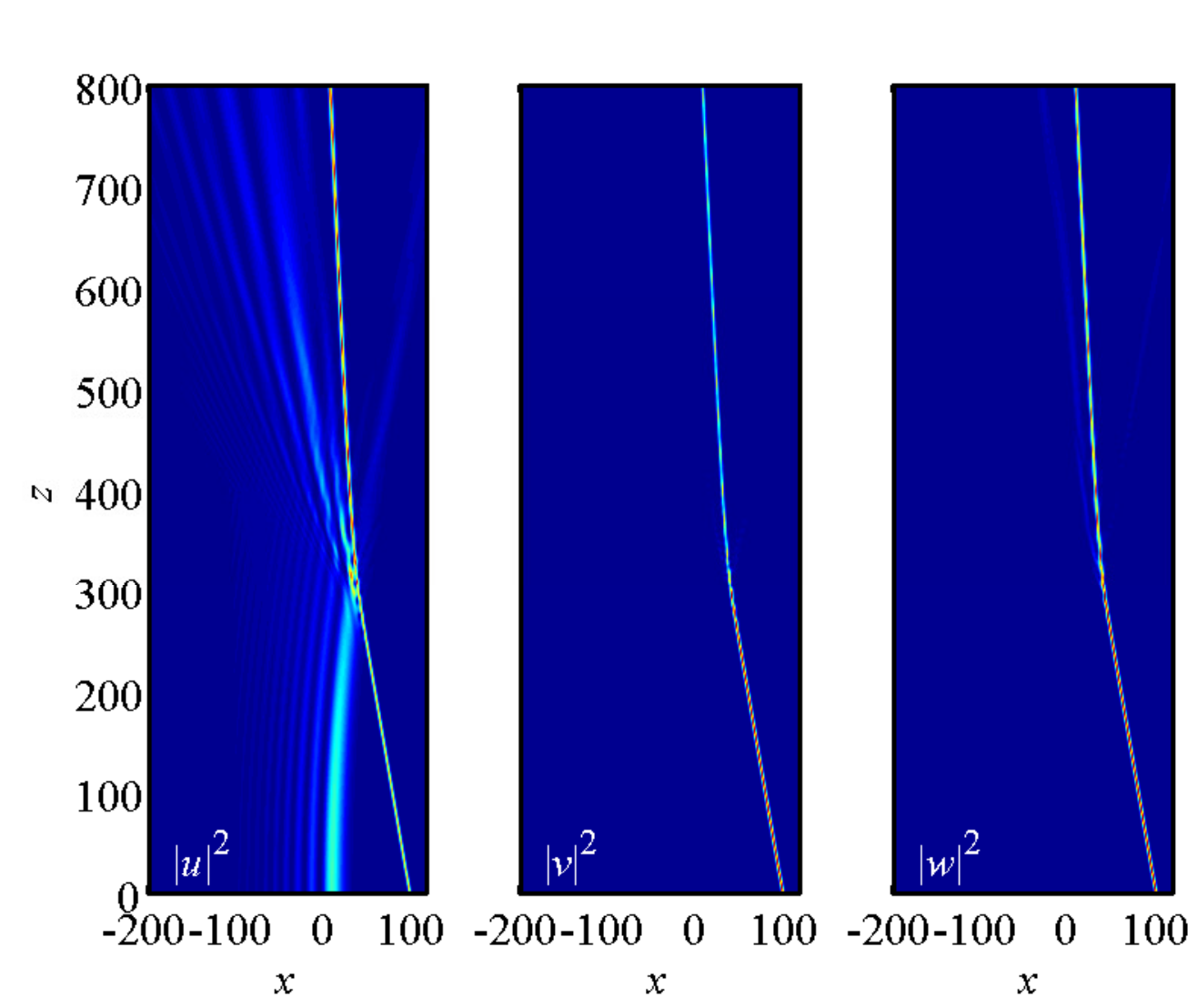}}%
\subfigure[]{\includegraphics[width=3in]{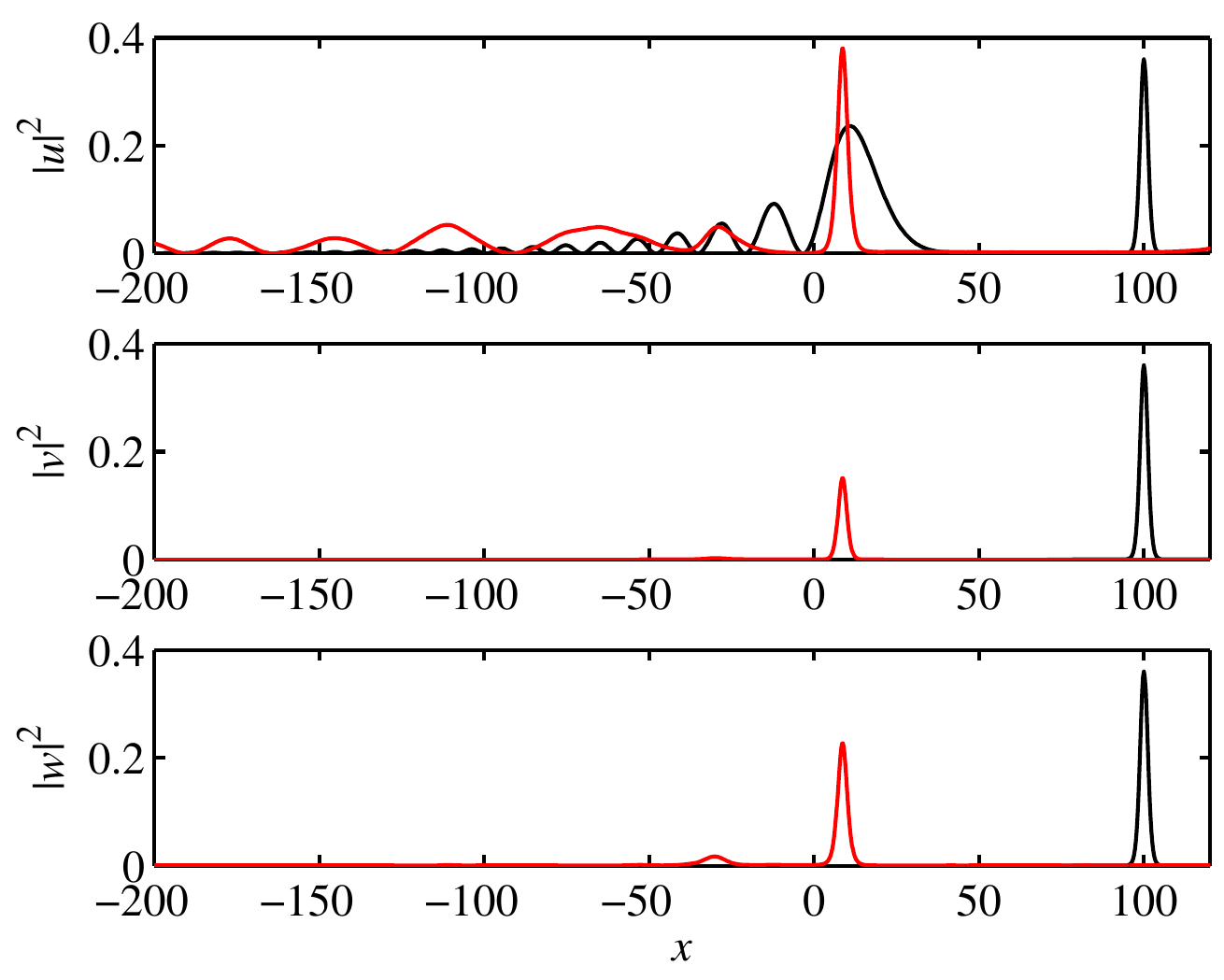}}
\caption{(Color online) The same as in Figs. \protect\ref{fig8}-\protect\ref%
{fig10}, but for $q=-1.20$. In this case, the collision between the TAW and
soliton with nearly equal total powers creates a bound complex including
both waveforms (in a distorted form), which has an almost zero velocity.}
\label{fig11}
\end{figure}

With the increase of $P_{\mathrm{sol}}$ to the value close to $P_{\mathrm{TAW%
}}$, which corresponds to $q=\allowbreak -1.20$ (in this case, Eq. (\ref%
{P1DAW}) yields $P_{\mathrm{sol}}=6.\,\allowbreak 44\approx 1.02$ $P_{%
\mathrm{TAW}}$), the simulations demonstrate the most complex outcome in
Fig. \ref{fig11}: the original TAW and symmetric soliton with nearly equal
powers merge into a complex which keeps both the Airy and soliton components
discernible, although the latter one becomes asymmetric, and the TAW is
essentially distorted too. The merger leads to cancellation of the TAW's and
soliton's velocities, so that the bound complex emerges in a nearly
quiescent form.

\begin{figure}[tbp]
\centering\subfigure[]{\includegraphics[width=3in]{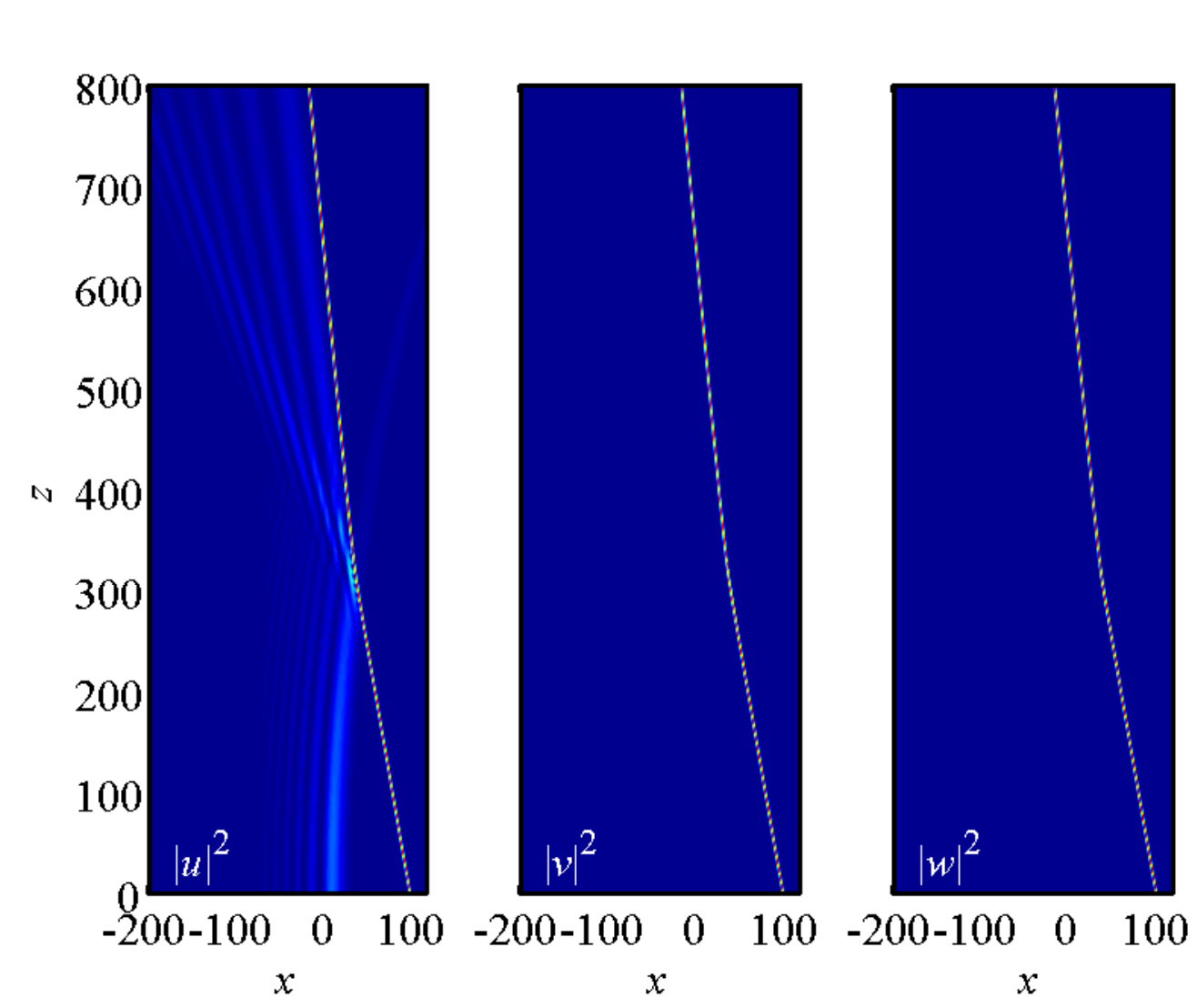}}%
\subfigure[]{\includegraphics[width=3in]{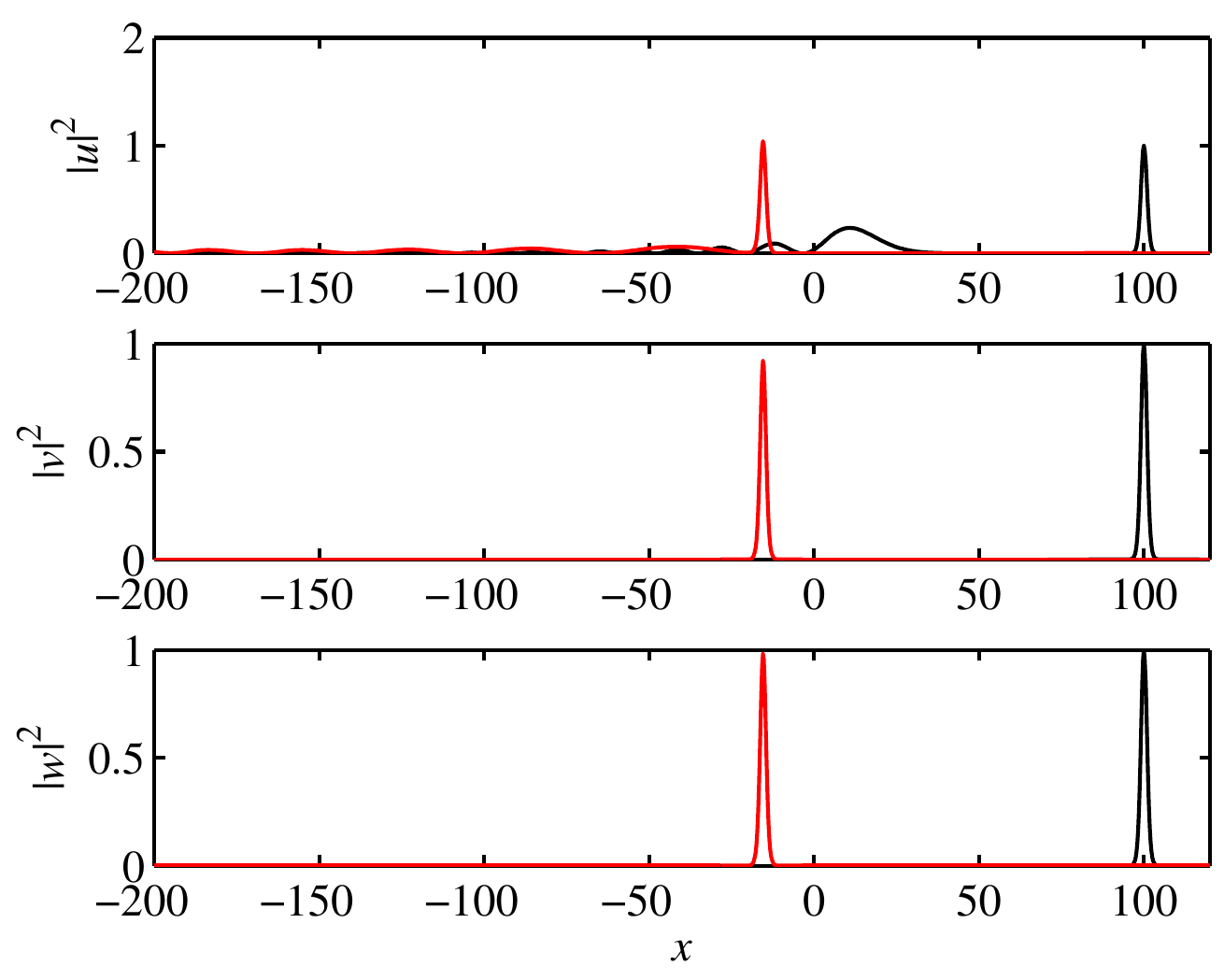}}
\caption{(Color online) The same as in Figs. \protect\ref{fig8}-\protect\ref%
{fig11}, but for $q=-2.0$. In this case, the heavy soliton absorbs the TAW
with which is collides.}
\label{fig_additional}
\end{figure}

Finally, when $P_{\mathrm{sol}}$ becomes essentially larger than $P_{\mathrm{%
TAW}}$, \textit{viz}., at $P_{\mathrm{sol}}\geq $ $\allowbreak
11.\,\allowbreak 71\approx \allowbreak 1.\,\allowbreak 86~P_{\mathrm{TAW}}$,
which corresponds to $-q\geq 1.9$, the collision leads to absorption of the
TAW by the high-power soliton. A typical example of this outcome is shown
for $-q=2.0$, which corresponds to $P_{\mathrm{sol}}\approx \allowbreak
2.01P_{\mathrm{TAW}}$, in Fig. \ref{fig_additional}.

In the above, the outcomes of the strong TAW-soliton interaction,
corresponding to the relatively small collision velocity, $c=-0.2$, have
been demonstrated. Next, keeping $q=-1.20$, which, as said above,
corresponds to $P_{\mathrm{TAW}}\approx P_{\mathrm{sol}}$, i.e., the most
complex outcome of the collision, we aim to present results obtained for
increasing $|c|$, as the change of the outcome with the variation of the
velocity is a relevant issue too. The same outcome as displayed in Fig. \ref%
{fig11} for $c=-0.2$, i.e., the formation of the bound Airy-soliton complex,
is observed at $|c|\leq 0.3$.

\begin{figure}[tbp]
\centering\subfigure[]{\includegraphics[width=3in]{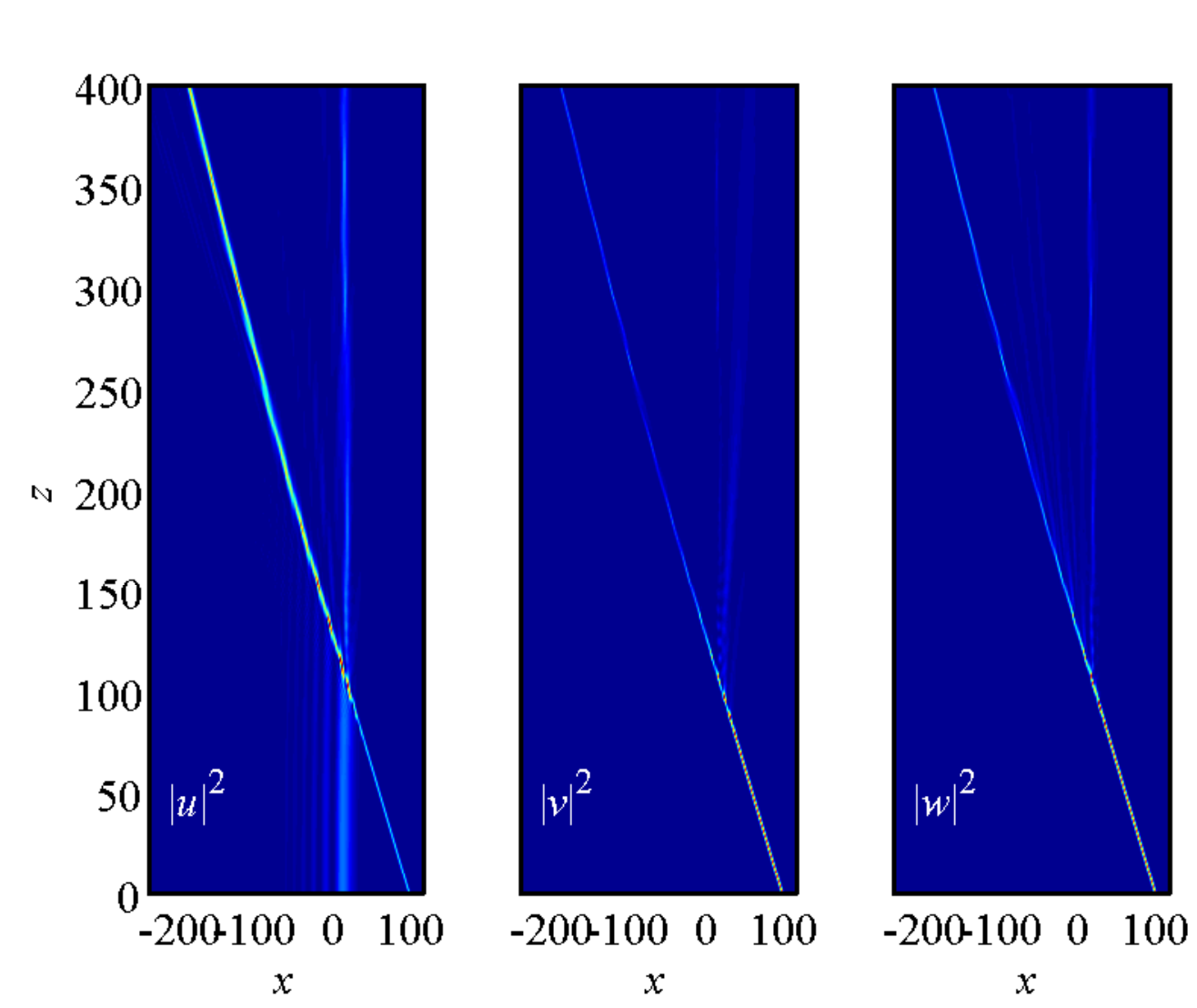}}%
\subfigure[]{\includegraphics[width=3in]{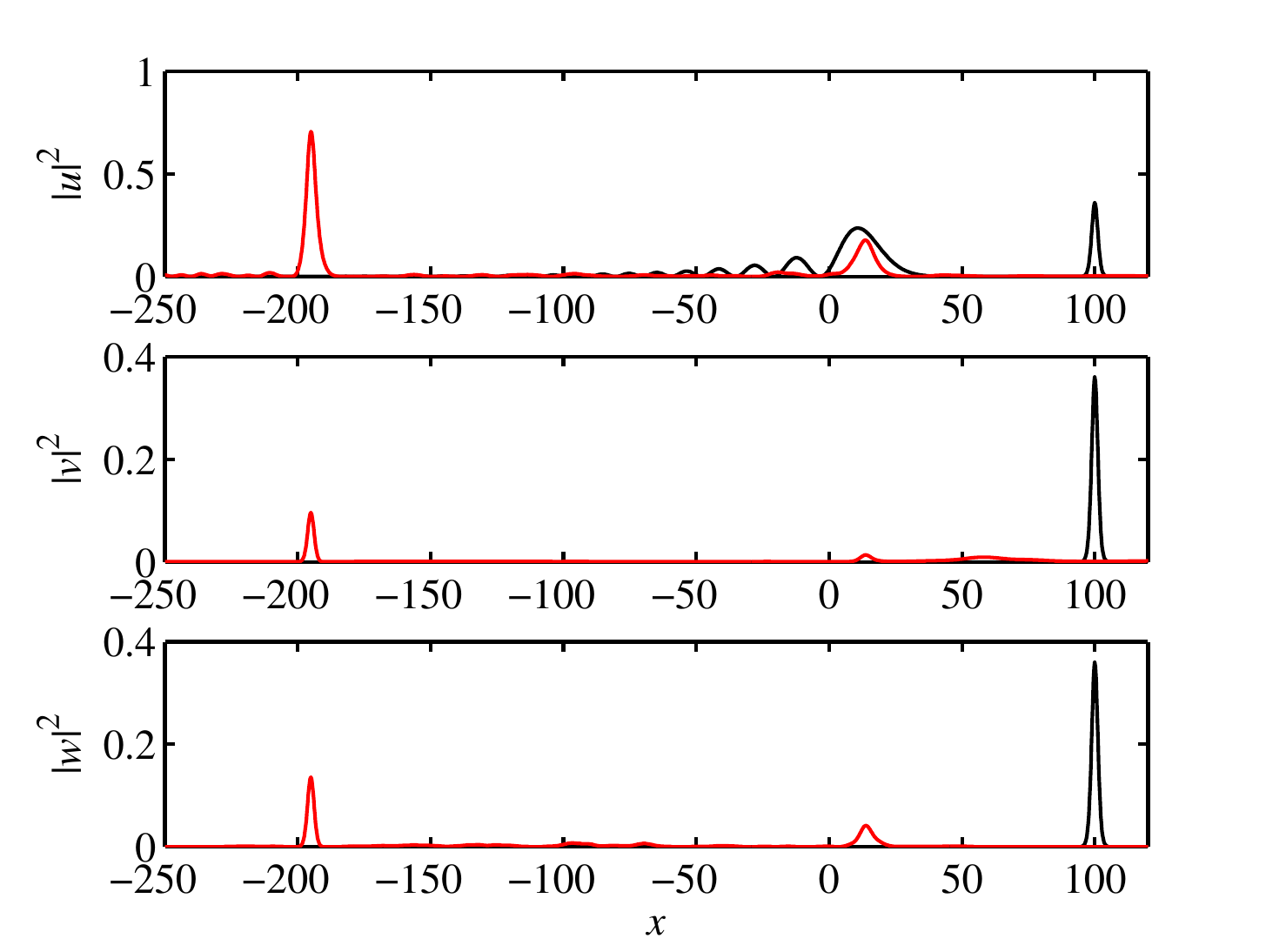}}
\caption{(Color online) The same as in Fig. \protect\ref{fig11}, but for the
incident soliton with velocity $c=-0.8$. In this case, the TAW-soliton
collision transforms the interacting modes into a relatively simple set of
two solitons, which continue their motion along the pre-collision
directions. }
\label{fig12}
\end{figure}

In the interval of velocities $0.3<|c|\leq 1.0$, the velocity is large
enough to allow separation of the modes after the collision. In this case,
Fig. \ref{fig12} demonstrates that the interaction, which is still strong
enough, leads to a simple result, similar to that displayed above in Fig.\ %
\ref{fig9} -- the transformation of the Airy-soliton pair into a set of two
asymmetric solitons, which follow the pre-collision directions of the
incident soliton and TAW. It is relevant to note that detailed consideration
of the data demonstrates that the increase of the power in the $u$-component
of the original soliton, observed in Fig. \ref{fig11}, is explained not by
snatching the power from the TAW, but by its transfer from the $v$- and $w$%
-components of the same soliton.

In the interval of $1.0<|c|\leq 1.5$, which is represented by Fig. \ref%
{fig13}, the attenuation of the interaction, caused by the increase of the
collision velocity, allows the TAW to partly keep its shape after the
collision. Further, at $1.5<|c|\leq 2.5$, the collision begins to feature
quasi-elasticity, although the incident soliton is still transformed from
the symmetric shape into a conspicuously asymmetric one, see a typical
example in Fig. \ref{fig14} for $c=-2.0$. In this case, like in the
situation displayed in Fig. \ref{fig12}, the increase of the $u$-component,
observed in its outcome profile, is explained by the transfer of power from
the soliton's $v$- and $w$-components, rather than by capturing power from
the original TAW.
\begin{figure}[tbp]
\centering\subfigure[]{\includegraphics[width=3in]{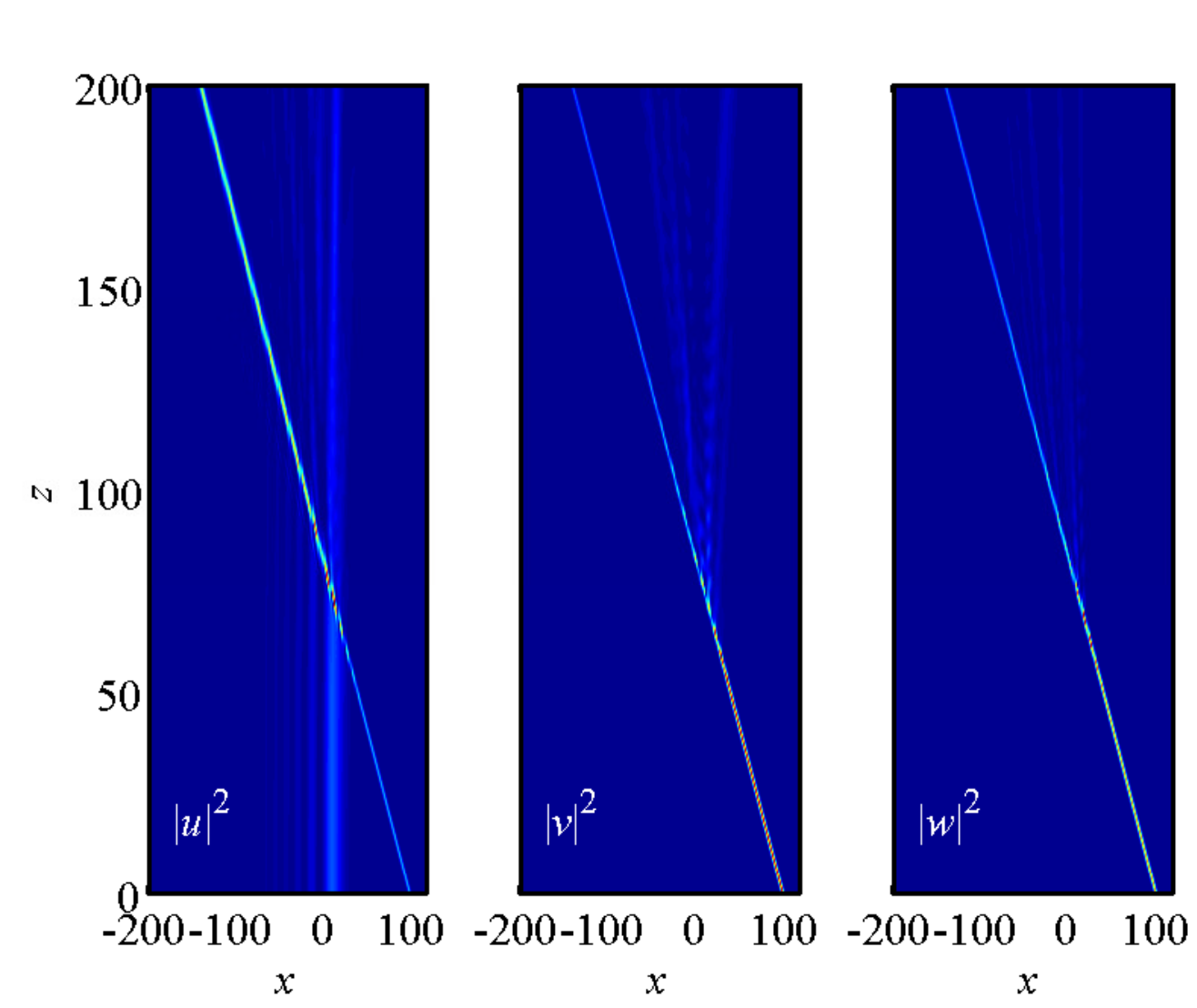}}%
\subfigure[]{\includegraphics[width=3in]{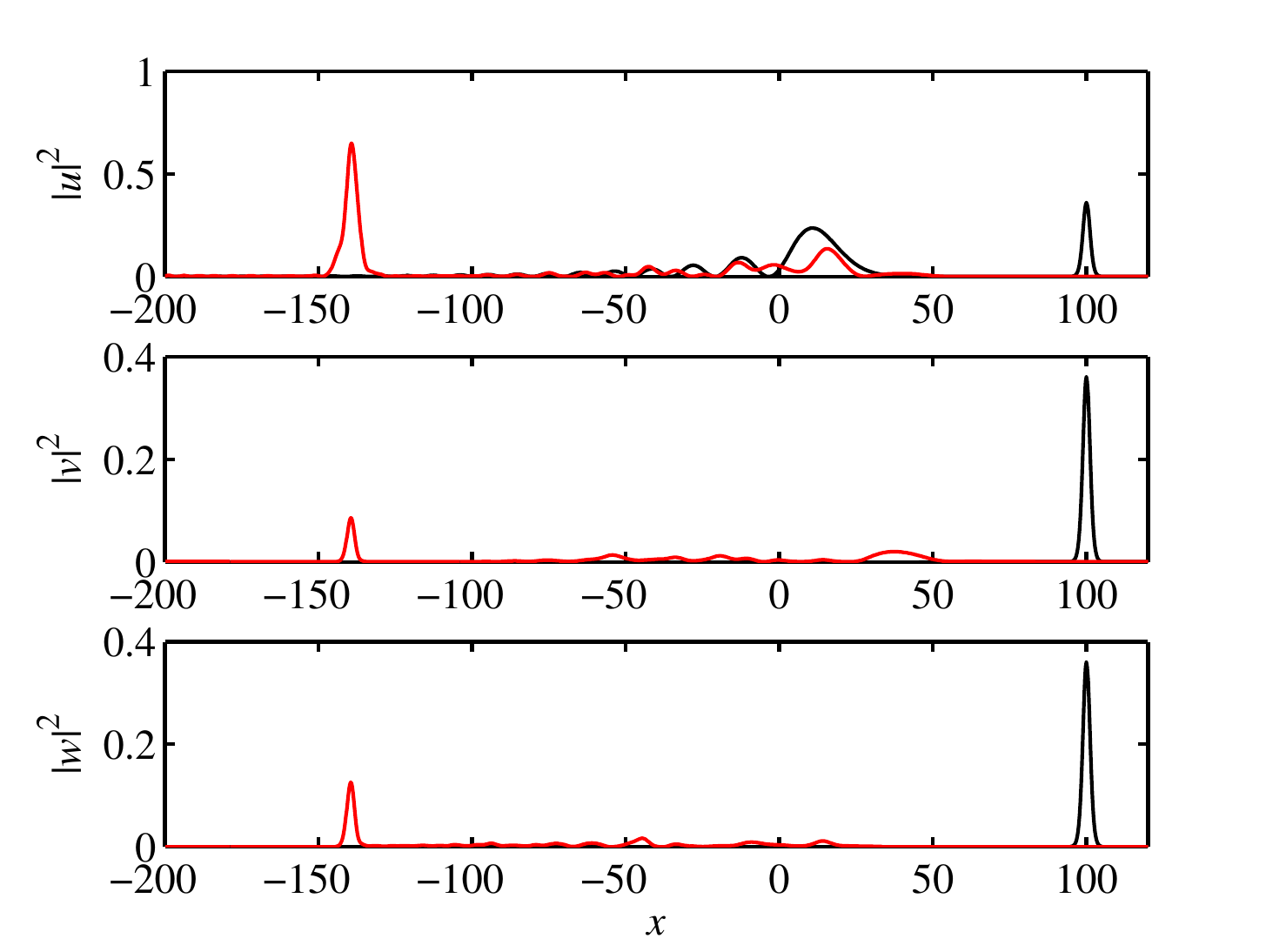}}
\caption{(Color online) The same as in Figs. \protect\ref{fig11} and \protect
\ref{fig12}, but for the incident soliton with velocity $c=-1.2$. In this
case, the TAW shape partly survives the collision.}
\label{fig13}
\end{figure}
\begin{figure}[tbp]
\centering\subfigure[]{\includegraphics[width=3in]{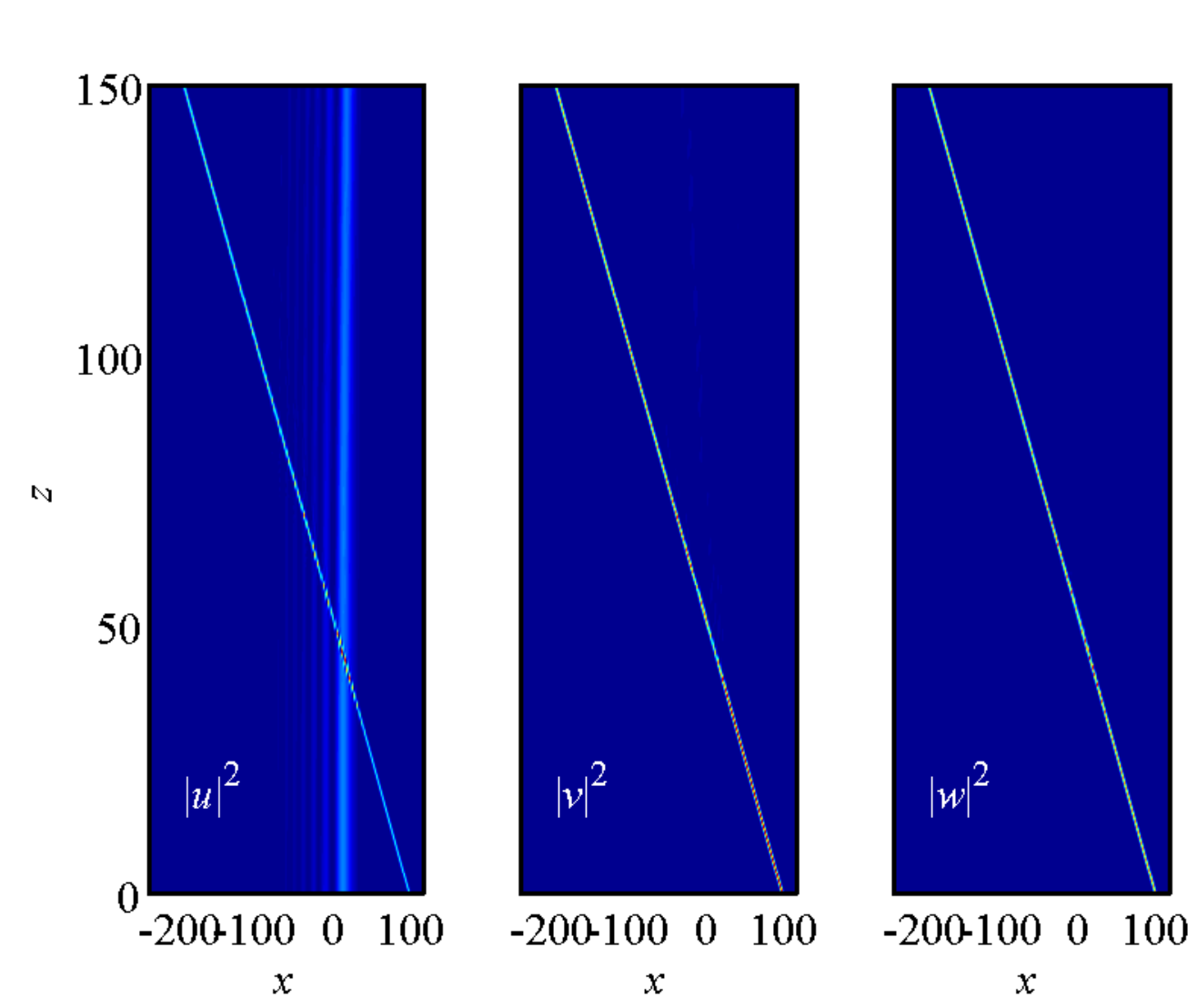}}%
\subfigure[]{\includegraphics[width=3in]{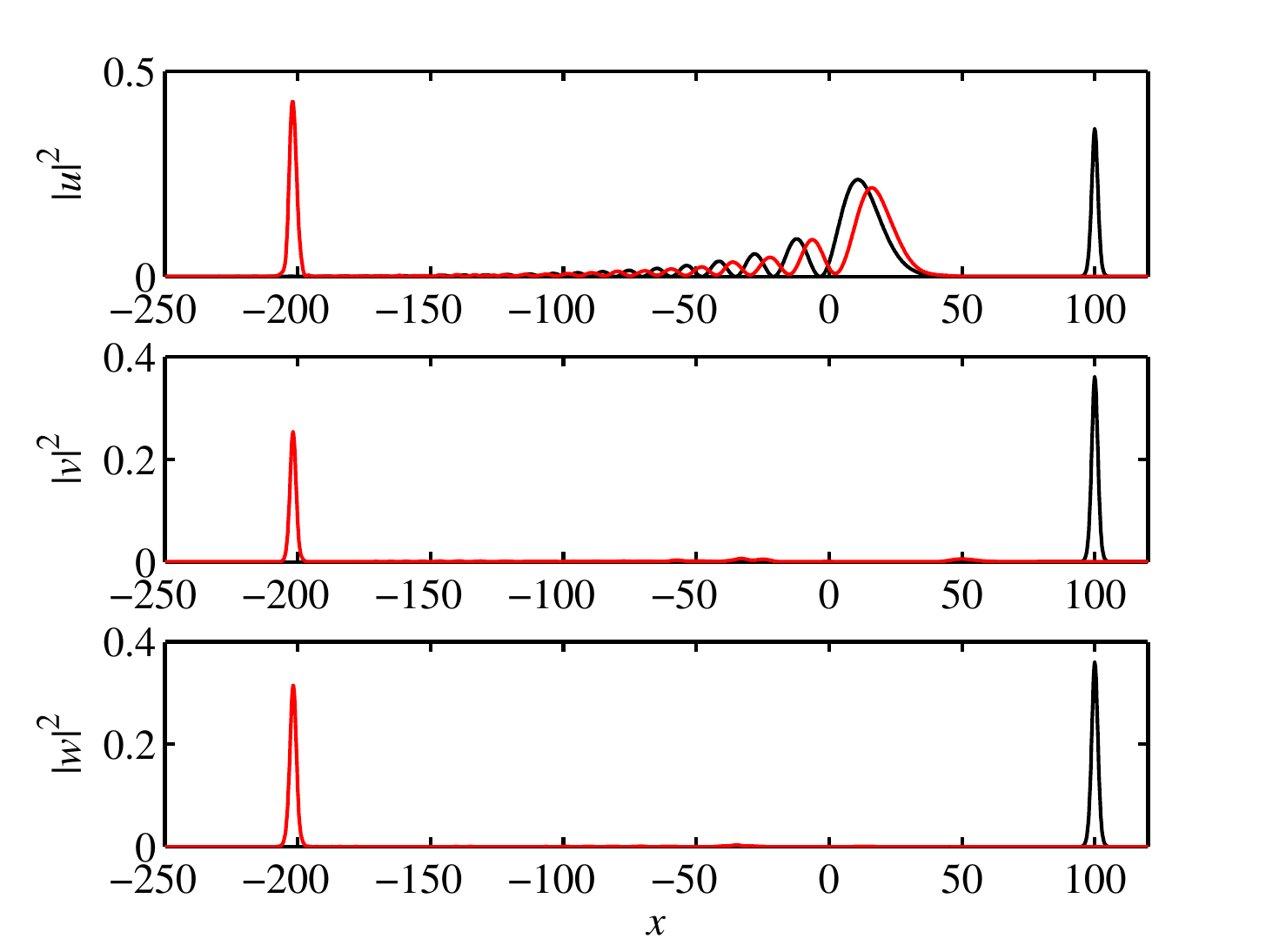}}
\caption{(Color online) The same as in Figs. \protect\ref{fig11}, \protect
\ref{fig12}, and \protect\ref{fig13}, but for the incident soliton with
velocity $c=-2.0$. In this case, the Airy-soliton collision begins to
feature quasi-elasticity.}
\label{fig14}
\end{figure}

\begin{figure}[tbp]
\centering\subfigure[]{\includegraphics[width=3in]{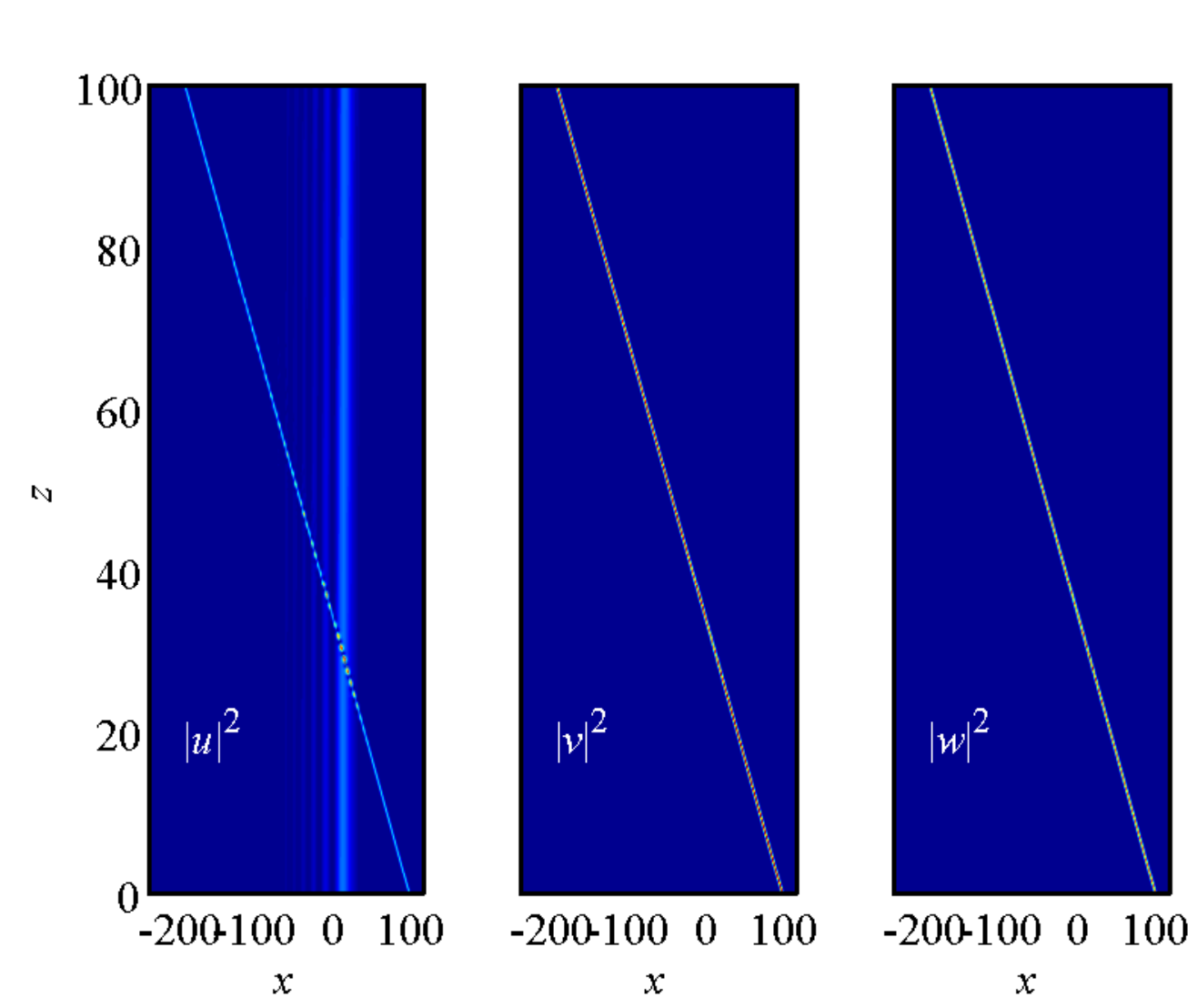}}%
\subfigure[]{\includegraphics[width=3in]{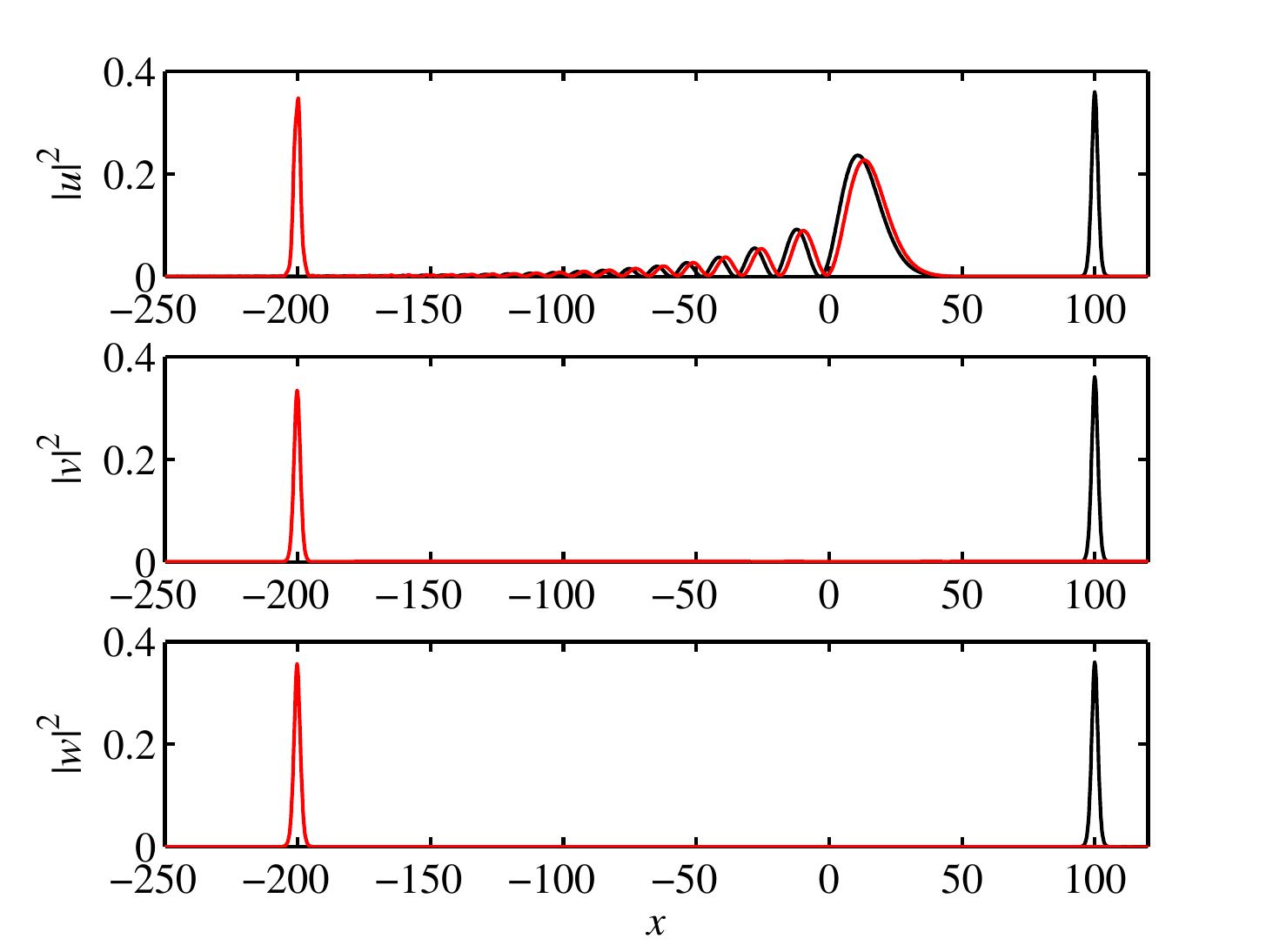}}
\caption{(Color online) The same as in Figs.\protect\ref{fig11} and \protect
\ref{fig12}-\protect\ref{fig14}, but for the incident soliton with velocity $%
c=-3.0$. In this case, the collision is fully elastic.}
\label{fig15}
\end{figure}

Lastly, the collision becomes fully elastic at $|c|>2.5$, as is clearly seen
in Fig. \ref{fig15} at $c=-3.0$. Note that the incident symmetric soliton
keeps its symmetry in this case.

While the above results were presented for the collisions of the TAW with
the moving symmetric solitons, given by exact solution based on Eqs. (\ref%
{k12}) and (\ref{exact}), collisions of the Airy wave with asymmetric
three-wave solitons, such as the one displayed in Fig. \ref{fig1}, have been
systematically simulated too (not shown here in detail). It has been
concluded that the results are not qualitatively different from those
summarized above for the collisions with the symmetric solitons.

\section{Conclusion}

The objective of this work is to demonstrate that the generic three-wave
system coupled by the $\chi ^{(2)}$ interaction makes it possible to
consider interactions of the TAWs (truncated Airy waves) and three-wave
solitons in the setting which is not available in other nonlinear systems.
The advantage of the present system is that the single-wave TAWs, carried by
one FF (fundamental-frequency) component, which are not distorted by the
nonlinearity, are stable against parametric perturbations (unlike the
previously considered setting, with the parametrically unstable TAW in the
SH (second-harmonic) component), and the three-wave solitons are stable as
well (some additional numerical and analytical results for these solitons
were reported here too). While the TAWs are stable, they are found to be
rather fragile against collisions with each other, and with three-wave
solitons. In particular, the collision between two mutually symmetric TAWs
carried by different FF components leads to the formation of a cluster of
solitons, which eventually coagulate into a set of three ones with the
increase of the total power. As concerns TAW-soliton collisions, the TAW
absorbs incident solitons with a very small power, and a high-power soliton
absorbs the TAW. In the intermediate region, the collision breaks the TAW
into two solitons (with a remnant TAW attached to one of them), or forms a
complex TAW-soliton bound state. The increase of the collision velocity
gradually makes the collision quasi-elastic.

As an extension of the present work, it may be quite interesting to develop
the analysis for the two-dimensional three-wave system, in which stable
single-wave TAWs and three-wave solitons are available too, cf. Ref. \cite%
{we2}.

\section{Acknowledgment}

This work was supported by the Thailand Research Fund through grant No.
RSA5780061.

\end{document}